\documentclass[a4paper,11pt]{article}
\usepackage[]{xcolor}
\pdfoutput=1
\usepackage{subfig}
\usepackage{fixltx2e}
\usepackage{jcappub} 
\usepackage{float}
\usepackage{cprotect}
\usepackage{hyperref}
\usepackage{comment}
\usepackage{ulem}

\newcommand{\myrefA}[1]{\hyperref[#1]{\textbf{A}}}
\newcommand{\myrefB}[1]{\hyperref[#1]{\textbf{B}}}

\usepackage[T1]{fontenc}
\usepackage{subfig}

\title{\boldmath Towards cosmology with Void Lensing: how to find voids sensitive to weak-lensing and numerically interpret them }

\author[a]{Renan Boschetti}
\author[a]{Pauline Vielzeuf}
\author[a]{Marie-Claude Cousinou}
\author[a]{Stephanie Escoffier}
\author[b]{and Eric Jullo}
\affiliation[a]{Aix-Marseille Université, CPPM, IPhU, Marseille, France}
\affiliation[b]{Aix Marseille Univ, CNRS, CNES, LAM, Marseille, France}

\affiliation[a]{\small}

\emailAdd{renan.boschetti@gmail.com}

\cprotect\abstract{In this work, we present a study of the void lensing signal or the excess surface mass density (ESMD) around cosmic voids. First, we propose a new void-finder algorithm that is designed to capture the ESMD around voids. We compare our algorithm applied to projected slices with the \verb|ZOBOV| void finder and find significantly deeper weak-lensing profiles for voids defined by our algorithm in the context of a realistic galaxy mock. Then we test the consistency between the measurements of the ESMD as measured through the shear of background galaxies and directly calculated through the dark matter density profiles of the same voids. We found inconsistencies for voids with diameter $\geq 100h^{-1}\mathrm{Mpc}$ along the line-of-sight, but the consistency holds for smaller voids, meaning that we are indeed probing the underlying dark matter field by measuring the shear around these voids. Moreover, we show that voids found in the projected slices, which are highly sensitive to lensing, are correlated to $3$D voids exhibiting intrinsic alignments between them. }


\begin{document}

\maketitle
\flushbottom

\section{Introduction}

Given the abundance of cosmological data coming from Large-Scale Structure (LSS) surveys such as Euclid \cite{euclid-laurejis2011} and DESI \cite{aghamousa2016desi}, one of the central challenges of this era is to know how to interpret and optimize the extraction of relevant physical information. This task passes through the identification of discrete tracers of LSS, each one tracing the underlying dark matter field in a particular way and therefore carrying information about the details of the underlying Dark Matter (DM) structures. 

One particular type of tracer of LSS 
is 
cosmic voids, which 
consists,
generically speaking, of large underdensities located in-between clusters, filaments and walls, dominating the volume of the Universe. These voids typically range from a few $h^{-1}\textrm{Mpc}$ up to $\simeq 100 h^{-1}\textrm{Mpc} $ in radius. Hence, they populate the LSS in a very distinct manner compared to galaxies or halos, as evidenced, for instance, by their negative linear bias \cite{chan2020measurement}. Thus, these structures carry complementary information to luminous tracer statistics. In fact, it has been shown that the density profiles 
of voids, as well as their abundance, are particularly sensitive to dark energy \cite{pisani2015counting}, massive neutrinos \cite{massara2015voids, schuster2019bias}, primordial non-gaussianities \cite{kamionkowski2009void} and modifications to general relativity \cite{Voivodic_2017, perico2019cosmic}. Arguably, since voids are underdense in DM, their evolution should be more sensitive to dark energy, modifications to gravity and neutrinos.

Despite their potential as a cosmological probe, using voids in cosmological analysis is challenging for a couple of reasons, amongst them are the high shot-noise in the case of void auto-correlation, the presence of cosmic variance and the mismatch between voids found in the sparse galaxy field and the true voids in the underlying DM field, as well as the lack of theoretical knowledge regarding the distribution of matter in the vicinity of the void center (see \cite{massara2018density} for a recent development in this direction). 

 One particular observable involving voids is their imprint on the shape of background galaxies, the so-called void lensing (VL). Unlike overdensities, around which the shapes of background galaxies are strengthened around the target structure, voids will leave the opposite imprint: galaxies shapes ellipticies tend to be radially orientated around the void, as a result of an ``anti-lensing'', or negative tangential shear. It has received increasing attention in the last two decades, since the first proposition (up to our knowledge) by \cite{amendola1999weak} in 1998. Since then, a few measurements were made \cite{sanchez2016cosmic, fang2019dark, melchior2014first}, as well as analytical and numerical investigations have shown the sensitivity of this observable to modifications of gravity \cite{barreira2015weak, baker2018void, davies2019cosmological}. 
 
Weak-lensing combined with voids can mitigate the limitation imposed by the sparsity of luminous tracers, since it is sensitive to the total matter and therefore we are able to ``see'' voids in this field. Furthermore, weak-lensing in general is useful to test gravity because it is sensitive to the sum of the Newtonian potentials $\Psi_{L}=(\Phi + \Psi)/2$ which in general are not equal in modified gravity scenarios. Basically, by measuring the distortion on the shapes of distant galaxies when their light passes through voids, we are directly (the total matter) probing the cleanest (less baryonic complications), simplest (less non-linearities) and more sensitive environment to modified gravity.

In the literature, the detection of weak lensing around voids is basically treated in two distinct ways: (i) by measuring the tangential shear $\gamma_{t}$, or the convergence $\kappa$, which parameterize the distortion on the shapes of background galaxies caused by the underdense structures spanning  from the observer up to the source \cite{shimasue2023line, gruen2016weak, higuchi2013measuring, davies2021optimal} and (ii) by measuring the Excess Surface Mass Density (ESMD), $\Delta \Sigma$, which is basically the projection of the total matter contained in a thin lens (in this case, voids) localized somewhere in-between the observer and the source. Therefore, the approach (i) is measuring the projected profile of ultra-large troughs (with sizes of hundreds of $h^{-1}\textrm{Mpc}$ ), whereas the approach (ii) is measuring the projected profiles of voids with radii $\leq 50 h^{-1} \textrm{Mpc}$, which is the limiting radius for which the thin lens approximation is still valid,  as we will show in this work. The aim of the approach (ii) is to extract dynamical and morphological information about the usual generic definition of voids, i.e., underdense structures located in-between overdense structures (halos, sheets and filaments). The inference of void profiles and mass function through lensing can potentially give cleaner information about voids in the total matter and be a complementary to void analysis by, for instance, being able to measure the void bias \cite{fang2019dark}, whereas approach (i) measures the weak-lensing signature of ultra-large structures along the line-of-sight, the so-called troughs, and the relation of these objects to a generic definition of voids (with predictable void abundance) is not clear. Furthermore, approach (i) has less 3D information, which approach (ii) partially recovers. Although approach (i)  could be interesting on its own, here we make a distinction between them  and call by ``void lensing'' approaches such as (ii) and by ``troughs'' the approach (i). This work investigates aspects of VL. 

 Previous works on void lensing used two definitions of voids \cite{fang2019dark, sanchez2016cosmic}. In short, the first definition finds voids in the $3$D galaxy field and the second finds voids in projections of the galaxy field with width of $100 h^{-1}\textrm{Mpc}$. The first approach has the advantage of having trivial interpretation, i.e., $\Delta \Sigma$ is basically the projection of the void profile, but has the disadvantage of signal contamination by overdensities surrounding the voids. The second approach has no clear interpretation in terms of void definitions in the 3D field, but has the potential of obtaining higher S/N per void, since these voids fill the whole projected slices along the line-of-sight, minimizing the contamination by overdensities.
 
In this work we aim to compare VL measurements for two types of void-finders: one which is widely used in literature and also in previous measurements of VL in real data, based on the \verb|ZOBOV| algorithm \cite{fang2019dark, sanchez2016cosmic}, and the one that we introduce in this work. We show that it is worth exploring the different sensitivity of these different approaches to cosmology and modified gravity, due to the very distinct profiles of voids that they yield. We also investigate the consistency between the VL signal as measured through shear, and the same signal as estimated directly using the DM field. If we aim to do precision cosmology with this observable, we have to, first of all, be sure that we really measure the projected void profile through the shear in a controlled environment. As we will show, when working with voids to measure the ESMD, problems related to the size of voids might arise. We believe this work paves the way for a numerical interpretation of void lensing and therefore for precision cosmological analysis using the data from up-coming surveys with large sky-fraction coverage.

 This work is organized as follows. In section \ref{sec:wl}, we briefly review the weak lensing basics. 
In section \ref{sec:pheno}, we describe the void-finding algorithm introduced in this work, as well as the resulting profile and abundance of voids measured in a simulation. In section \ref{sec:meas_cmp}, we compare the performance of our void-finding algorithm with the widely used \verb|ZOBOV| void finder by measuring the ESMD in a galaxy mock. In section \ref{sec:num_interp}, we compare the ESMD as measured through the tangential shear $\gamma_t$ and the one measured directly from the DM field of the same realisation, giving then a numerical interpretation of the observational ESMD and addressing the issue of the thin-lens approximation in the context of void lensing.   Finally, we conclude with some observational considerations and avenues for future works.

\section{ Weak-Lensing theory}
\label{sec:wl}

In this section, we briefly review a few results of the weak-lensing theory that constitute the basis of this work. We refer to \cite{schneider2005gravitational} for an extensive review.

As the light of background galaxies propagate through LSS, its path is perturbed by the gravitational field in the foreground. The difference between the unperturbed and observed positions, respectively $\boldsymbol{\beta}$ and $\boldsymbol{\theta}$, is given by the lens equation: 
\begin{equation}
    \boldsymbol{\beta} = \boldsymbol{\theta} - \boldsymbol{\alpha}\,.
\end{equation}
\noindent Assuming small scalar perturbations to the Minkowsky metric, $\boldsymbol{\alpha }$ is given by the projection of the gradient of the gravitational potential along the line of sight:

\begin{equation}
    \boldsymbol{\alpha} = \frac{2}{ c^{2}}\int^{\chi_{S}}_{0} d\chi'\frac{(\chi_{S} - \chi'  )\chi'}{\chi_{S}}\boldsymbol{\nabla}_{\perp}\Phi(\boldsymbol{\theta}\chi', \chi').
\end{equation}

 The integral is accounting for all lenses at positions $\chi'$ up to the source position $\chi_{S}$. The factor $(\chi_{S} - \chi)\chi/\chi_{S}$ is a geometrical weight of the projection. 
Therefore, it is expected that the observed shapes of background galaxies will be different than the original shapes. The distortion in those shapes is expressed by the distortion matrix

\begin{equation}
    A_{ij} = \frac{\partial \beta_{i}}{\partial \theta_{j}} =\delta_{ij} - \frac{\partial \boldsymbol{\alpha}}{\partial \theta_{j}} = \delta_{ij} - \frac{\partial^{2} \psi }{\partial \theta_{i} \partial \theta_{j}},
    \label{dist_matrix}
\end{equation}
\noindent which can be written as 
\begin{equation}
\boldsymbol{A} =
\begin{pmatrix}
  1 - \kappa - \gamma_{1} & -\gamma_{2} \\
 -\gamma_{2} & 1 - \kappa + \gamma_{1}
       
\end{pmatrix}.
\label{a_mat}
\end{equation}

In eq. \ref{dist_matrix} we have defined the lensing potential

\begin{equation}
\psi = \frac{2}{c^{2}}\int^{\chi_{S}}_{0} d\chi' \frac{(\chi_{S} - \chi')}{\chi_{S}\chi'}\Phi (\boldsymbol{\theta}\chi', \chi').
\end{equation}

The convergence, $\kappa$, is related to the increase or decrease of the overall image size, whereas $\gamma_{1,2}$ parameterize the deviation of the image from a circle and are given by $\gamma_{1} = 1/2(\partial_{1}\partial_{1} - \partial_{2}\partial_{2})\psi$, $\gamma_{2} = \partial_{1}\partial_{2}\psi$. Using the previous results, the convergence ($\kappa = 1/2(\partial^{2}_{1} + \partial^{2}_{2})\psi$) can be written as

\begin{equation}
    \kappa = \frac{1}{4 \pi G} \int^{\chi_{s}}_{0} \frac{\nabla^2 \Phi }{\Sigma_c} \mathrm{d} \chi,
\label{conv_def}
\end{equation}

\noindent where $\Sigma_{c} = c^{2}\chi_{S}/ (4\pi G \chi (\chi_{S} - \chi) )$ is the critical surface mass density. The tangential ($E$ mode) and cross ($B$ mode) components of the shear are defined respectively as $\gamma_{t} = -\mathbb{R}[\gamma e^{-2i\phi}]$ and $\gamma_{\times} = -\mathbb{I}[\gamma e^{-2i\phi}]$, where $\gamma = \gamma_{1} + i\gamma_{2}$. Notice that by factorizing out the convergence in Eq.~\ref{a_mat}, the deviation from identity becomes the reduced shear $g = \gamma/(1 - \kappa)$, which is the actual observable. In the weak field regime (our case), $g \simeq \gamma $ is a reasonable approximation. The tangential shear $\gamma_{t}$ is positive in the case in which the foreground lenses are overdensities and negative in the case of underdensities, whereas the cross-component, $\gamma_{\times}$, is related to curl, which is not produced by weak-lensing and therefore must vanish.

In the case of axially symetric lenses, we can write the tangential shear in terms of the convergence as 

\begin{equation}
\gamma_{t}(r_{\perp}) = \bar{\kappa}(< r_{\perp}) - \kappa(r_{\perp}), 
\end{equation}
\noindent where 

\begin{equation}
\bar{\kappa}( < r_\perp) = \frac{2}{r_{\perp}^{2}}\int^{r_\perp}_{0} r_{\perp}' \kappa (r_\perp') dr_\perp.
\end{equation}

In this work, we are interested in measuring the projected profile of voids in the thin lens approximation. Therefore, we take $\Sigma_{c}$ out of the integral in Eq.~\ref{conv_def}, and integrate only in the radial bin that we regard as acting as a thin lens. By inserting the Poisson equation $
\boldsymbol{\nabla}^{2} \Phi(\boldsymbol{\theta}\chi, \chi, a) = 4 \pi G a^{2}\bar{\rho}_{m}(a)\delta(\boldsymbol{\theta}\chi, \chi)$, the convergence then becomes

\begin{equation}
   \kappa = \frac{3 H^{2}_{0} \Omega^{(0)}_{m}}{8 \pi G\Sigma_c} \int^{\chi_{l} + \Delta \chi/2}_{\chi_{l} - \Delta \chi/2}  \frac{\delta(\boldsymbol{\theta}\chi, \chi)}{a(\chi)} \mathrm{d} \chi,
   \label{tl_def}
\end{equation}

\noindent where $\Delta \chi$ is the bin width in comoving distance, which we consider as a thin lens. 
 Based on this approximation, we define the ESMD in terms of the tangential shear, which is the quantity that we can measure in photometric surveys :

\begin{equation}
\Delta \Sigma(r_{\perp}) = \Sigma_{c}\gamma_{t}(r_\perp).
\label{delta_sig_def}
\end{equation}

On the other hand, we can directly calculate the ESMD as

\begin{equation}
    \Delta \Sigma (r_{\perp}) = \bar{\Sigma}(<r_{\perp}) - \Sigma(r_{\perp}), 
    \label{DS_def_proj}
\end{equation}
\noindent where

\begin{equation}
    \Sigma(r_{\perp}) = \frac{3 H^{2}_{0} \Omega^{(0)}_{m}}{8 \pi G} \int^{\chi_{l} + \Delta \chi/2}_{\chi_{l} - \Delta \chi/2}  \frac{\delta_{v}(\boldsymbol{\theta}\chi, \chi)}{a(\chi)} \mathrm{d} \chi
    \label{Sigma_def}
\end{equation}

\noindent and

\begin{equation}\label{eq:sigbar}
    \bar{\Sigma}(<r_{\perp}) = \frac{2}{r^{2}_{\perp}}\int^{r_{\perp}}_{0}dr'_{\perp} r'_{\perp}\Sigma(r'_{\perp}).
\end{equation}

In the context of this work we use the void density contrast, $\delta_{v}$, to calculate $\Delta \Sigma$. 

\section{Optimum Centering Void Finder (OCVF)}
\label{sec:pheno}

In this section we present the void finder algorithm we developed for this work. First we give the intuition and the recipe. Then we show two void statistics produced by this algorithm applied to a N-body simulation: the void density profile and the void abundance. 

\subsection{Intuition and recipe}
\label{sec:pheno_profile}

In order to have an intuition for how a void finder algorithm should be to deliver an ideal VL profile \footnote{henceforth, we use VL profile and ESMD ($\Delta \Sigma$) interchangeably}, we explore the possibilities of VL signals produced by voids described by an hyperbolical tangent-like profile: 

\begin{equation}
    \delta_{v}(r|R_{v}) \equiv \frac{\rho_{v}\left(r \mid R_{v}\right)}{\bar{\rho}_{m}} - 1 = |\delta_{c}|\left\{   \frac{1}{2}\left[1+\tanh \left(\frac{y-y_{0}}{s\left(R_{v}\right)}\right)\right] -1 \right\} ,
    \label{ht_profile}
\end{equation}
\noindent where $y = \ln(r/R_{v})$, $y_0= \ln(r_{0}/R_{v})$, $R_{v}$ is the void radius and $r_{0} = 0.37s^{2} + 0.25s + 0.89$, which is calibrated to describe voids with average density which is $\bar{\rho}_{v}(< R_{v})/\bar{\rho}_{m} = 0.2$ of the average density of the Universe, where 
\begin{equation}
\bar{\rho}_{v}(<R_{v}) = \frac{3}{r^{3}_{v}}\int^{R_{v}}_{0}dr r^{3} \bar{\rho}_{m}(\delta_{v}(r) + 1),
\end{equation}
$s$ parameterises the gradient of the profile and $\delta_{c}$ the density contrast at the void center. This profile was first introduced in \cite{voivodic2020halo}. 

We choose to use this profile instead of the widely used HSW profile \cite{hamaus2014universal} because the latter has more free parameters and therefore would make this exercise more complicated. However, it is important to notice that the HSW is more general than the profile \ref{ht_profile}, which is particularly suitable in our case.

 This exercise will give us an intuition for what kind of voids we have to pursue in order to maximize certain desirable properties, namely the signal amplitude and emptiness, since emptier regions are more sensitive to the kind of physics we are interested in when working with voids, namely modifications to gravity, neutrino masses, or complementary information from underdensities which will increase the constraining power in a multi-tracer analysis.

 Comparisons between different types of void finders was made in the context of N-body simulations by \cite{cautun2018santiago}, where they found that voids found in the projected DM field have more power of distinguishing between modified gravity and GR. Also, \cite{davies2021optimal} shows that the same type of voids provide the highest S/N. However, the void finder (called Tunnels) which presents the desirable features in both works have overlapping between the voids. Our algorithm is intended to have similar voids but without overlapping, which is desirable in order to be able to predict the theoretical void abundance \cite{sheth2004hierarchy}.

\begin{figure}
\centering
\begin{minipage}[b]{.4\textwidth}
\includegraphics[width=7cm]{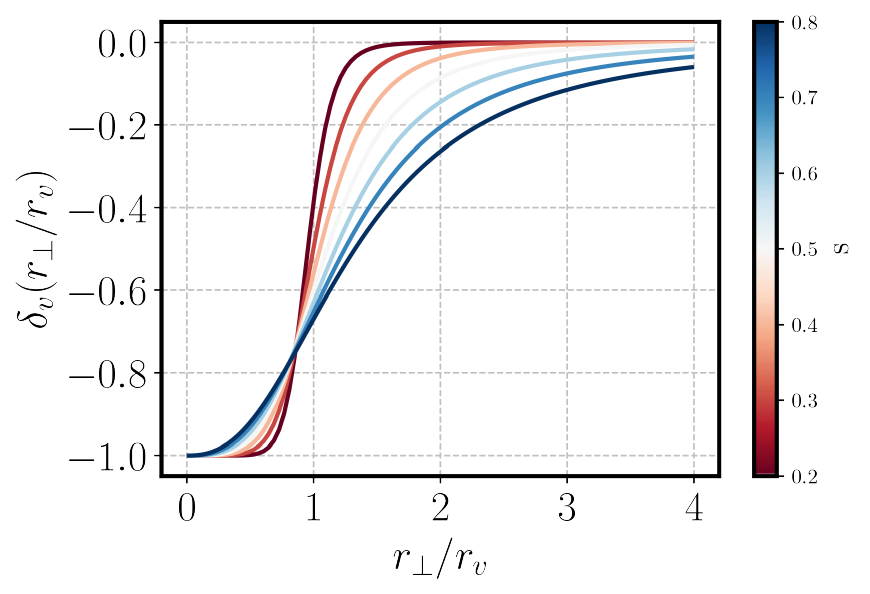} 
\end{minipage}\qquad
\begin{minipage}[b]{.4\textwidth}
 \includegraphics[width=7cm]{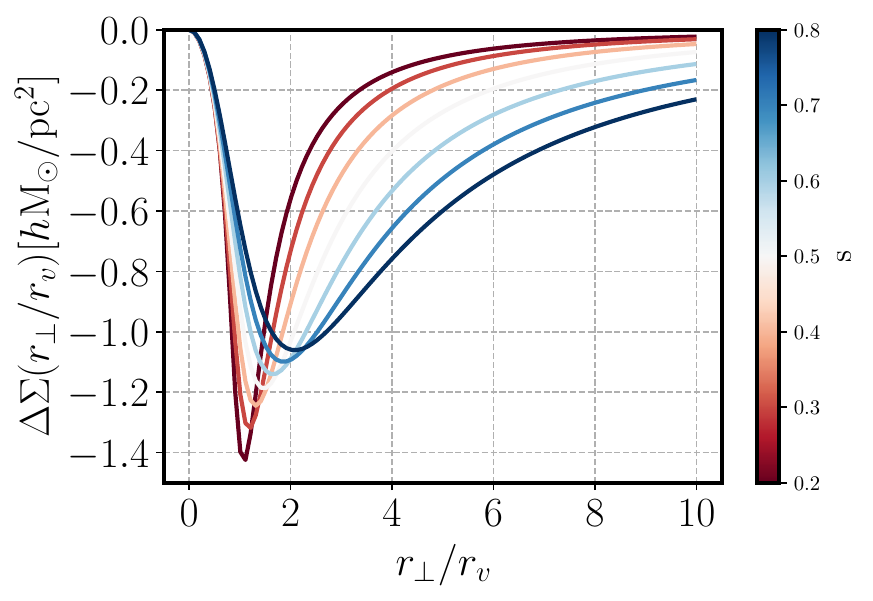} 
\end{minipage}
\caption{Left: Generic void profiles of type eq. \ref{ht_profile} for different choices of $s$, which controls the profile gradient. Right: the resulting differential surface densities, $\Delta\Sigma$ obtained through eq. \ref{DS_def_proj}. This exercise shows that deeper $\Delta \Sigma$ profiles are produced by voids with density profiles which have drastic transitions between $\delta_c$ and the average density, $\bar{\rho}_{m}$. }
\label{var_s}
\end{figure}

\begin{figure}
\centering
\begin{minipage}[b]{.4\textwidth}
\includegraphics[width=7cm]{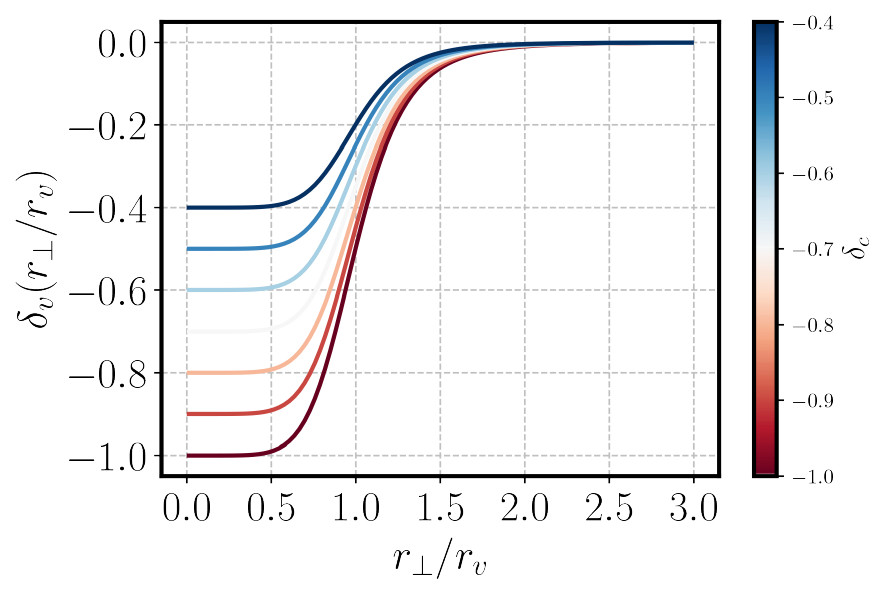} 
\end{minipage}\qquad
\begin{minipage}[b]{.4\textwidth}
 \includegraphics[width=7cm]{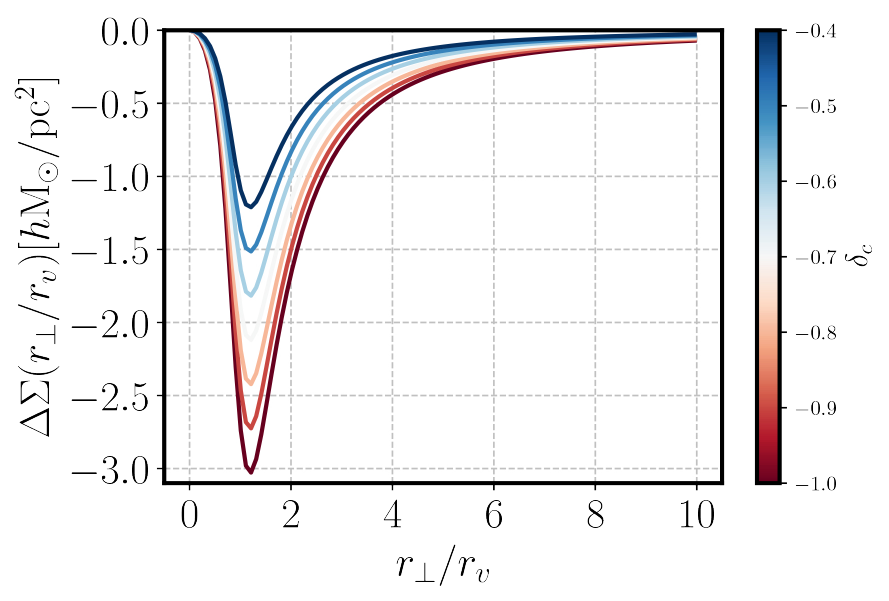} 
\end{minipage}
\caption{Left: Generic void profiles of type eq. \ref{ht_profile} for different choices of $\delta_{c}$, which controls the density at the void center position. Right: the resulting differential surface densities, $\Delta\Sigma$ obtained through eq. \ref{DS_def_proj}. This exercise shows that voids with smaller values of density at its center, $\delta_{c}$ produces deeper differential surface densities, $\Delta \Sigma$.}
\label{var_delta_c}
\end{figure}

Figures \ref{var_s} and \ref{var_delta_c} show, respectively, the void density profiles (left) and the corresponding VL profile (right) using Eq. \ref{DS_def_proj}, when one  varies the parameter $s$ (with fixed $\delta_{c}$) and the density at the void center $\delta_c$ (with fixed $s$) in the hyperbolical tangent profile ( Eq. \ref{ht_profile}). By 
Fig. \ref{var_s} it is clear that voids with profiles that go faster from 
their minimum density to the average density $\bar{\rho}_{m}$ (smaller $s$) produce deeper VL profiles, with minimum at the void radius, which means that these voids are emptier inside. Figure \ref{var_delta_c} shows that voids less dense in their centers also produce deeper profiles. 

Therefore the void finder must satisfy two criteria for defining each void in the catalogue: (i) 
the voids must have the smallest possible value of density at the center, $\delta_{c}$ (ii) 
their central position must be as far as possible from overdensities. The criteria (i) can be satisfied by the usage of Delaunay triangulation to define void center candidates.

The Delaunay triangulation is a set of d-symplexes (triangules for d = 2 or thetrahedrons for d = 3) $\mathbb{D}(\boldsymbol{P})$ defined in a set of discrete points $\boldsymbol{P}$, such that no point in $\boldsymbol{P}$ is inside any cirum-hypersphere of $\mathbb{D}(\boldsymbol{P})$. In the case d = 2,  the Delaunay triangles are those which are circumscribed by circles devoided of any point in $\boldsymbol{P}$ in their interior. In our context, the discrete set $\boldsymbol{P}$ can be any discrete tracer of LSS. By defining void positions as the centers of circum-hyperspheres, or circles circumscribed in Delaunay triangles in d = 2, we guarantee that criteria (i) is being satisfied. 

The criteria (ii) means that it is not enough to define void positions in empty regions, but also that amongst all the void position candidates we must choose the one which is further away from overdensities. This idea is based on the intuition that voids are empty regions surrounded by walls, filaments and clusters, all of which are overdense structures. Therefore, there is a point for each underdense region which must be further away from these structures. Having all the candidates from the Delaunay triangulation, the point we are looking for is the one from which we can grow the largest possible circle (for d = 2) until it reaches a certain fraction of the average density of the Universe, $\bar{\rho}_{v}(<R_{v})$, which is a free parameter of the void finder. Since all the candidates will have the same average density defined by $\bar{\rho}_{v}(<R_{v})$, the largest is the one which satisfies criteria (ii). 

Hence the algorithm can be roughly expressed in three steps:

\begin{itemize}
\item Perform the Delaunay triangulation to obtain the set of points which are candidates as void positions

\item Grow circles (or spheres) around them until the average density of these circles reach a certain density threshold, specified by $\bar{\rho}_{v}(<R_{v})$.

\item The largest circle will be the first void in the catalogue and all the other voids which intercept it will be discarded. The same process will be repeated for the second largest remaining void. This process will be repeated until a void which has radius smaller than the cutting radius $R_{c}$ is included in the catalogue.

\end{itemize}

The value of the cutting radius $R_{c}$ is arbitrary and can be regarded as a free-parameter. 

This algorithm also captures the ``essence'' of the excursion set theory \cite{sheth2004hierarchy}, since it is a way of finding the voids which correspond to the trajectories which first cross the threshold for void formation.

\subsection{Void profile and abundance}

In this section, we present the resulting void profiles and abundance after applying the algorithm described in the last section to an N-body simulation. The simulation we use in this section is a $1h^{-1}\textrm{Gpc}$ DM only box from the MultiDark suite \cite{prada2012halo}. The simulation has $3840^{3}$ DM particles which we sub sampled to have 
a density of 1 $h^{3}\textrm{Mpc}^{-3}$. The cosmology in this simulation is $(h, \Omega_{\Lambda}, \Omega_{m}, \Omega_b, n, \sigma_{8}) = (0.67, 0.69, 0. 307, 0.048, 0.96, 0.82)$.

\subsubsection*{Void Profile}
\begin{figure}[h!]
\centering
  \includegraphics[scale=0.45]{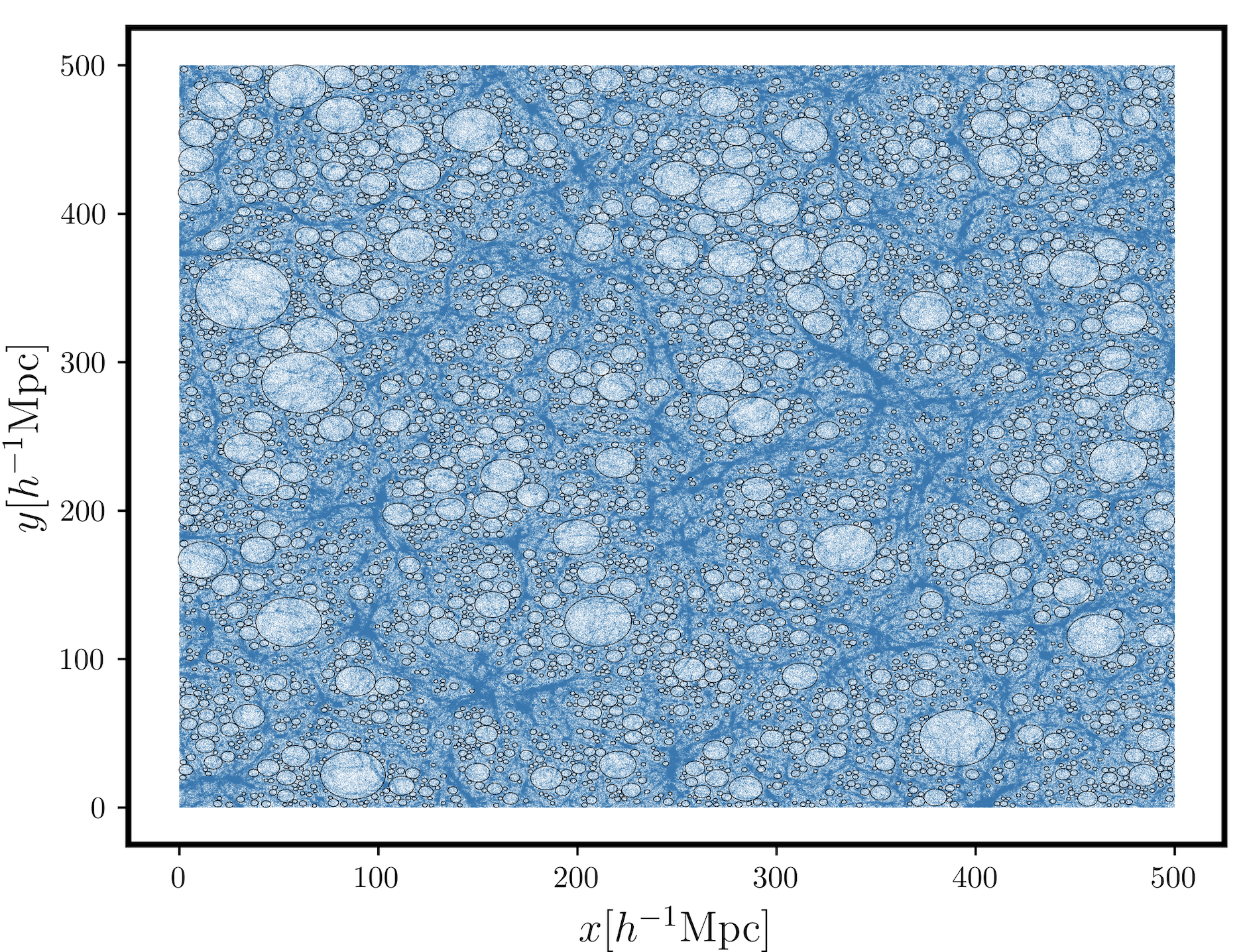}
  \caption{Visualisation of voids found in the projection of a slice of 50 $h^{-1}$Mpc of DM particles. }
  \label{voids_sim_MD}
\end{figure}

 We apply the algorithm to the simulation in its 3D version. For the purpose of visualisation, we show in 
 Fig. \ref{voids_sim_MD} the voids found in a 2D slice of 50 $h^{-1}\mathrm{Mpc}$. It is clear that voids are well fitted in underdense regions surrounded by filaments and walls, specially the largest voids. Although we don't show the same feature in the 3D field, we expect the same result, since the algorithm is exactly the same. 

\begin{figure}
\centering
\includegraphics[width=10cm]{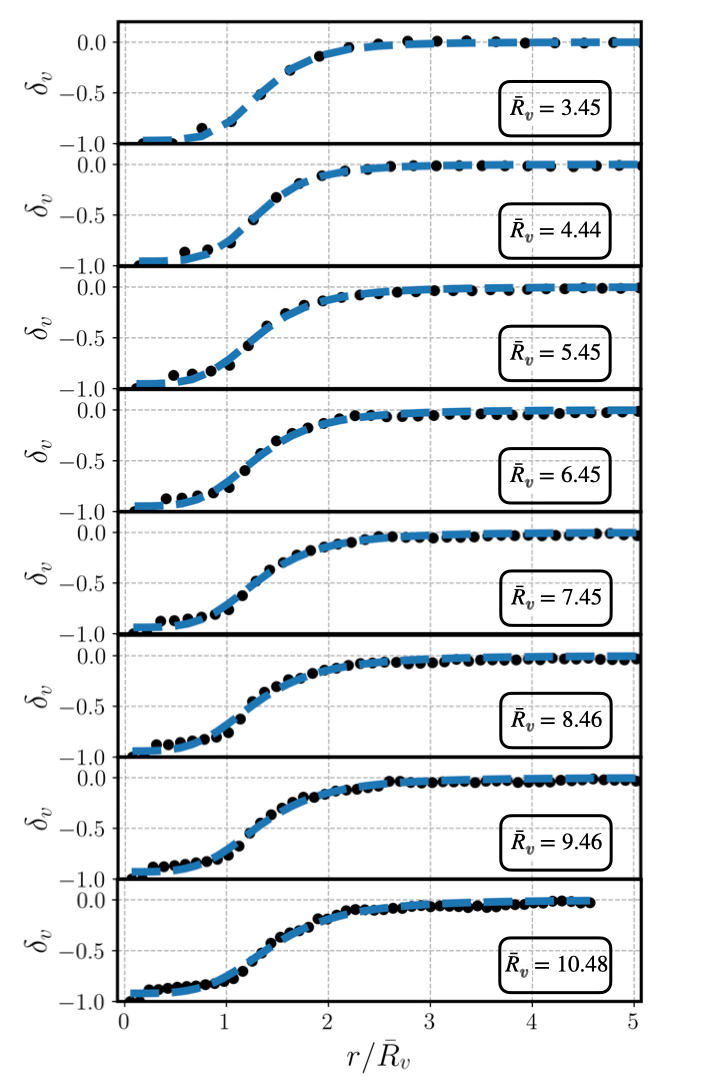} 
\caption{The measured 3D void profiles, $\delta_{v} = \rho_{v}/\bar{\rho}_{m}-1$ for each bin of radius compared to the profile given by Eq. \ref{ht_profile} \cite{voivodic2020halo}. The radii are given in units of $h^{-1}\mathrm{Mpc}$.}
\label{Profiles_sim}
\end{figure}

We can understand how voids are 
on average by looking 
at their stacked profiles, i.e  $\delta_{v}(r') = \rho_{v}(r')/\bar{\rho}_{m}-1$, where $r' \equiv r/R_{v}$ and $\rho_{v}$ is the density averaged over a spherical shell at reduced distance $r'$ to the void center. Figure \ref{Profiles_sim} shows the measured void profiles (black dots), compared to the fit profile given by Eq.~\ref{ht_profile} (dashed-blue). We fit the free parameter $s$ in each bin of radius $R_{v} = \{[3,4], [4,5], [5,6], [6,7], [7,8], [8, 9], [9, 10], [10,12]\}$ obtaining $s = \{0.38, 0.4, 0.46, 0.48, 0.51, 0.52, 0.53, 0.54\}$, meaning that the 
smallest voids 
goes 
 from the minimum density to $\delta_c = -1$ slightly faster than large voids. The reason for this is that smaller voids are mainly found inside overdense regions, being ``voids-in-clouds'' \cite{sheth2004hierarchy}, they don't present internal structure but rather they are simply almost empty places in the process of being collapsed by the overdense surroundings. In contrast, the largest voids present more substructures (this can be clearly seen in 
Fig. \ref{voids_sim_MD}), smoothing out their profiles.

The density profiles do not present a compensation wall such as the \verb|ZOBOV| voids \cite{neyrinck2008zobov}. This is directly related to the usage of the Delaunay approach that defines void center positions in empty places, and the \verb|OCVF| post-processing, where only the largest void in every region is kept and all the intercepting ones are discarded. This approach selects voids better placed into the ``holes'' in LSS. Consequently, the averaged density profile will not present the compensation wall - a signature of overdensities nearby void center positions. It is important to notice that these voids will not be more useful than the \verb|ZOBOV| ones, but will simply be different tracers of LSS.

\subsubsection*{Void Abundance}

The void abundance (void mass function, or void radius function) is the void analog of mass function for halos, i.e., the counts of objects per bin of mass. 

Several works have demonstrated its usefulness as a cosmological probe. For instance, \cite{perico2019cosmic} and \cite{cai2015testing} show the sensitivity of the abundance to modifications of gravity, \cite{kreisch2019massive} and \cite{massara2015voids} to massive neutrinos and \cite{verza2019void} to alternative dark energy scenarios. Recently, the first cosmological constraints using the void abundance was obtained by \cite{contarini2023cosmological}. 

The theory involved in the prediction of the void abundance is analogous to the theory developed for the halo mass function, which goes back to the pioneering work of Press-Schechter \cite{press1974formation} and its further developments (e.g. \cite{bond1991excursion, peacock1990alternatives, maggiore2010halo}). In the following, we briefly review the theory behind the prediction of the halo mass function. For a complete review, see reference \cite{desjacques2018large}. 

The so-called excursion set theory is based on the spherical collapse (expansion for voids), in which an isolated overdensity (underdensity) embedded into an Einstein de-Sitter background evolves and eventually collapses and virializes for halos, or experiences shell-crossing between internal shells which expand faster than outer shells for voids. Then, the linearly extrapolated value (from the initial conditions) of the density contrast for which the virialization (shell crossing) occurs is used as a threshold to define a halo (void). These values are $\delta_c = 1.686$ and $\delta_{v} = -2.7$, 
for halos and voids. 

In the excursion set, the Lagrangian density field is smoothed at some scale $R$ as

\begin{equation}
S(R) \equiv \sigma^2(R)=\left\langle\delta^2(x, R)\right\rangle=\int \frac{d^3 \boldsymbol{k}}{(2 \pi)^3} P_L(k) W_R^2(k),
\end{equation}
where $P_{L}(k)$ is the linear power spectrum and $W_{R}(k)$ is a smoothing function. The smoothed field $\delta(S)$ is then taken to perform a random walk, starting from $S = 0$ $(R\rightarrow \infty)$. The excursion set predicts the fraction of walks which will cross the threshold $\delta_c$ for the first time in the bin of mass $[M, M + dM]$, $f(M)$. This fraction is then converted into the number density of objects per bin of mass as

\begin{equation}
\frac{dn}{d \ln M} \equiv \frac{d^{2}N}{dV d\ln M} = \bar{\rho}_{m}f(M). 
\end{equation}
 
The exact form of the function $f(M)$ depends on assumptions regarding the smoothing function $W_{R}(k)$. The most popular choice, which simplifies the calculations, is a sharp-$k$ function \cite{bond1991excursion}. For this choice, the random walk performed by $\delta(S)$ is Markovian and then the probability that a walk first crosses the barrier $\delta_{c}$ is easily obtained as a solution of the Fokker-Planck equation with an appropriate boundary condition \cite{maggiore2010halo}. In this case, the halo mass function is given by
\begin{equation}
\frac{dn}{d \ln M} = \frac{\bar{\rho}_{m}}{M} f_{h}(\nu) \frac{d \ln \sigma^{-1}}{d \ln M}, 
\end{equation}
where $\nu = \delta_{c}/\sigma$, and the \textit{multiplicity function} for halos $f_{h}$  is defined as (in the Press-Schechter formalism)

\begin{equation}
f_{h}(\nu) = \sqrt{\frac{2}{\pi}} \nu e^{-\nu^{2}/2} \,.
\end{equation}

The prediction for the void abundance is analogous and was first proposed by \cite{sheth2004hierarchy}. The main difference from the halo case is the need for two density barriers, $\delta_{c}$ and $\delta_{v}$. So the problem reduces to finding the fraction of walkers which first cross $\delta_{v}$, without having never crossed $\delta_{c}$ for smaller $S$ (larger $R$). 

The solution obtained in \cite{sheth2004hierarchy} is

\begin{equation}
    \frac{d n_v}{d \ln R_{L}}=\frac{f^{2SB}_v(\sigma)}{V(R_{L})} \frac{d \ln \sigma^{-1}}{d \ln R_{L}},
    \label{void_abun}
\end{equation}
with the void multiplicity function given by
\begin{equation}
f_{v}^{2SB}(\sigma) =2 \sum_{n=1}^{\infty} \frac{n \pi}{\delta_T^2} S \sin \left(\frac{n \pi \delta_v}{\delta_T}\right) 
\exp \left(-\frac{n^2 \pi^2}{2 \delta_T^2} S\right).
\end{equation}

In the above equations $\delta_{T} = |\delta_{v}| + \delta_{c}$, the subscript $L$ and the superscript $2\mathrm{SB}$ stands for
, respectively, 
the linear radius and two static barriers. This radius is the linear comoving radius of the encompassing region in the Lagrangian space, which will expand until the shell crossing (non-linear radius). At the shell crossing, the void has expanded by a factor of $\simeq 1.7$ (see \cite{blumenthal1992largest}). The two static barriers refer to the density thresholds used in the excursion set, which are constant lines $\delta(S) = \delta_{c}, \delta_{v}$.

In reference \cite{jennings2013abundance}, it was noticed that the cumulative fraction of the number of voids exceeds unity. This was interpreted as a brake in the void number conservation. The solution given by Jennings et al. \cite{jennings2013abundance} was to impose the conservation of the volume fraction:
\begin{equation}
V(r)dn = V(r_{L})dn_{L}|_{r_{L}(r)}.
\end{equation}
where the linear and non-linear radii are related as $r\simeq 1.7 r_{L}$. Then the void abundance becomes
\begin{equation}
    \frac{d n_v}{d \ln R}=\frac{f^{2SB}_v(\sigma)}{V(R)} \frac{d \ln \sigma^{-1}}{d \ln R_{L}}.
    \label{void_abun_vdn}
\end{equation}

In the second paper of the series \cite{maggiore2010halo}, the static barrier for halo formation was generalised to a stochastic barrier. The stochastic barrier captures the arbitrariness involved in the halo/void finders, as well as complications which arise from environmental conditions. These might act against or in favour of halo/void formation and hence the critical density $\delta_{c}$/$\delta_{v}$ varies depending on the local variance. 

In \cite{sheth2004hierarchy}, an extended excursion set was first presented, i.e. the prediction for the abundance using two linear diffusing barriers:

\begin{equation}
\begin{aligned}
f_v^{2 \mathrm{LDB}}(\sigma) & =2\left(1+D_v\right) \exp \left[-\frac{\beta_v^2 \sigma^2}{2\left(1+D_v\right)}-\frac{\beta_v \delta_c}{1+D_v}\right] \\
& \times \sum_n \frac{n \pi}{\delta_T^2} \sigma^2 \sin \left(\frac{n \pi \delta_c}{\delta_T}\right) \exp \left[-\frac{n^2 \pi^2\left(1+D_v\right)}{2 \delta_T^2} \sigma^2\right].
\end{aligned}
\label{abund_pred}
\end{equation}

\begin{figure}[ht]
\centering
  \includegraphics[scale=0.8]{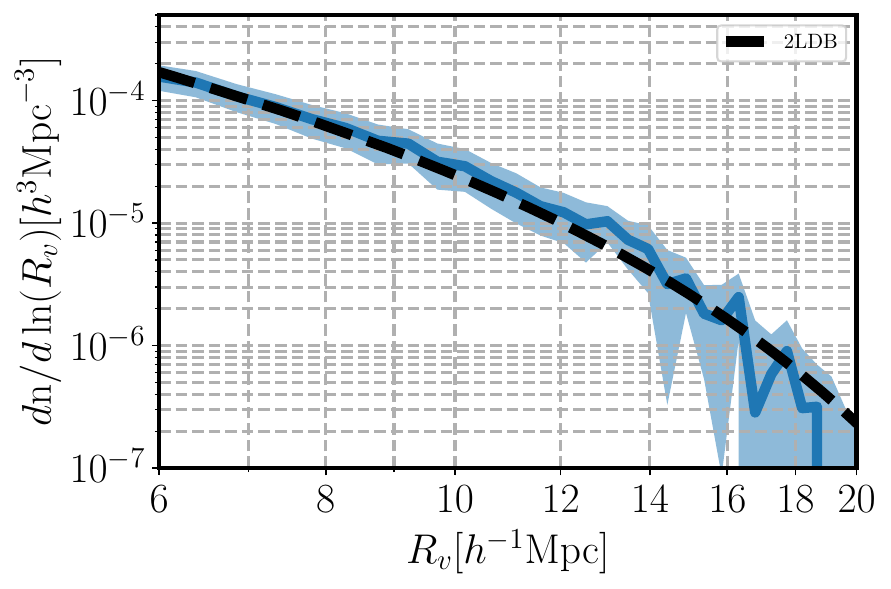}
\caption{The measured abundance in a DM only N-body simulation of size $L = 500h^{-1}\mathrm{Mpc}$ compared to the $2\mathrm{LDB}$ model prediction using two free-parameters ($\beta_{v}$, $D_{v}$).}
\label{abund_sim}
\end{figure}

 The superscript in $ 2\mathrm{LDB}$ stands for two linear diffuse barriers and ($\beta_{v}$, $D_v$) are free parameters which describe, respectively, 
 the slope and the diffusive coefficient of the barriers (see \cite{Voivodic_2017} for more details). Notice that we use the same slope and diffusive coefficient for the two barriers. The values of the linearly extrapolates thresholds for halo and void formation are kept fixed in $(\delta_{c} = 1.686, \delta_{v} = -2.7)$. 
 
 Figure \ref{abund_sim} shows the agreement between the abundance measured in the DM only simulation with size $L = 500h^{-1}\mathrm{Mpc}$ and the prediction given by the 2LDB prediction, with best fit free-parameters ($\beta_{v} = 0.02, D_{v} = 0.2$). The error bars are estimated by splitting the 1Gpc in 8 subboxes, estimating the abundance in each subbox and performing a JK in these 8 samples. The measured abundance is compatible with the theoretical expectation with only 2 free-parameters. 

\section{Void lensing on galaxies}
\label{sec:meas_cmp}

In this section we compare our algorithm \verb|OCVF| to  \verb|ZOBOV| (ZOnes Bordering On Voidness  algorithm) \cite{neyrinck2008zobov} in the context of void lensing. This algorithm is applied through the wrapper \verb|Revolver| \cite{nadathur2019revolver}. 

We apply both algorithms to the \verb|Buzzard| mock \cite{derose2019buzzard, derose2022dark, wechsler2022addgals, derose2022modeling}, which models the observed spectroscopic redshifts of galaxies matching a DESI-like survey (DeRose et al. (in prep)). The simulated light-cone covers $10,313.25$ $deg^{2}$, and contains 5,434,414 (BGS) 
galaxies brighter than $r=20.2$ distributed over the redshift range $z \in [0.1, 0.3]$. These galaxies are used as lenses, i.e., we find the voids using them as tracers. The source galaxy catalogue is modelled to match the photometric redshifts of a DES-like survey, occupying the same surface area, with density of 4.4 galaxies/arcmin$^{2}$ in the redshift range $z \in [0.5, 1.5]$. 

The main difference between the \verb|ZOBOV| and \verb|OCVF| is that the latter is Delaunay based, whereas the former is Voronoi based. Moreover, here we use the $2$D version of \verb|OCVF| and \verb|ZOBOV| finds voids in $3$D.  

\begin{figure}[h!]
\centering
\begin{center}
\centering
\includegraphics[width=13.cm]{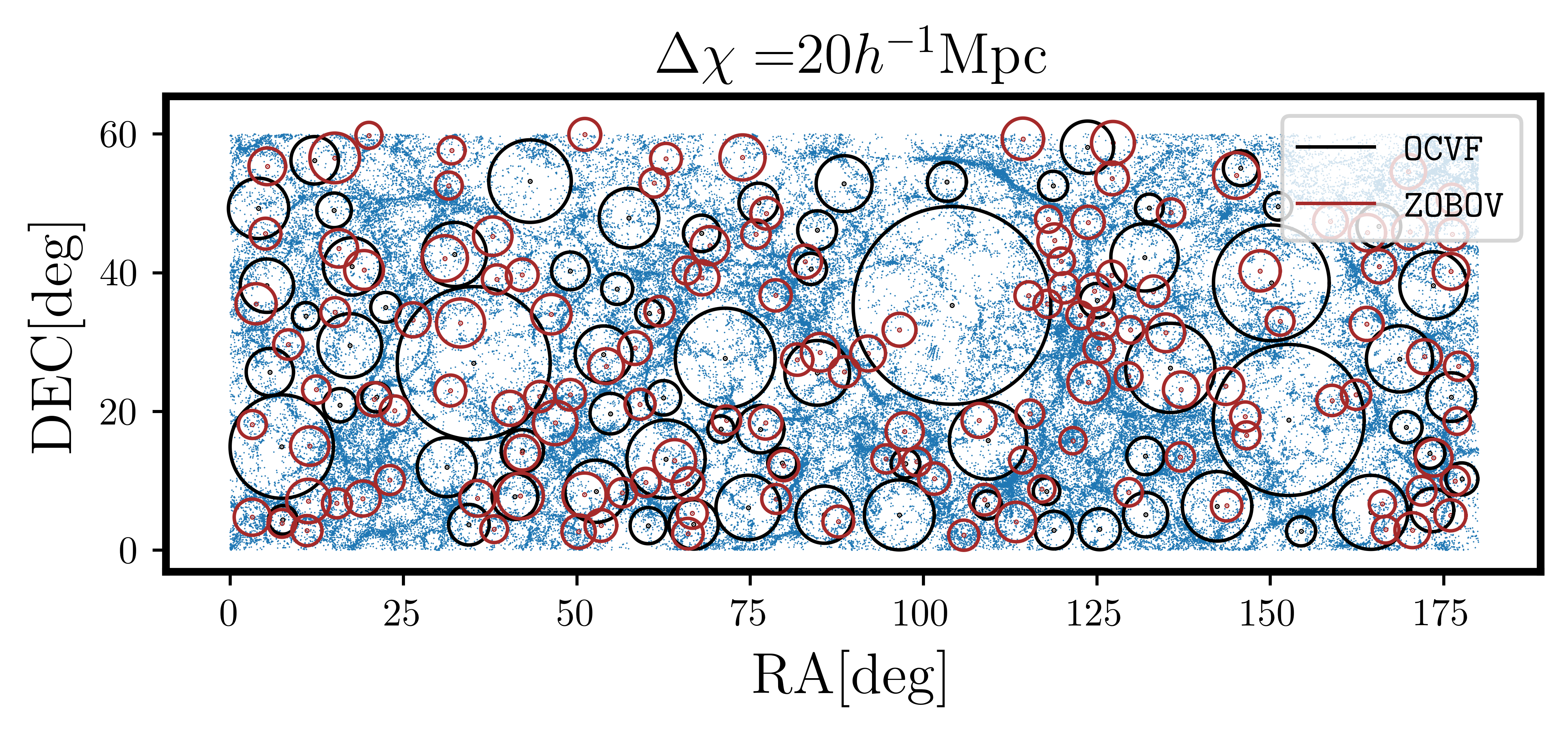}
\includegraphics[width=13.cm]{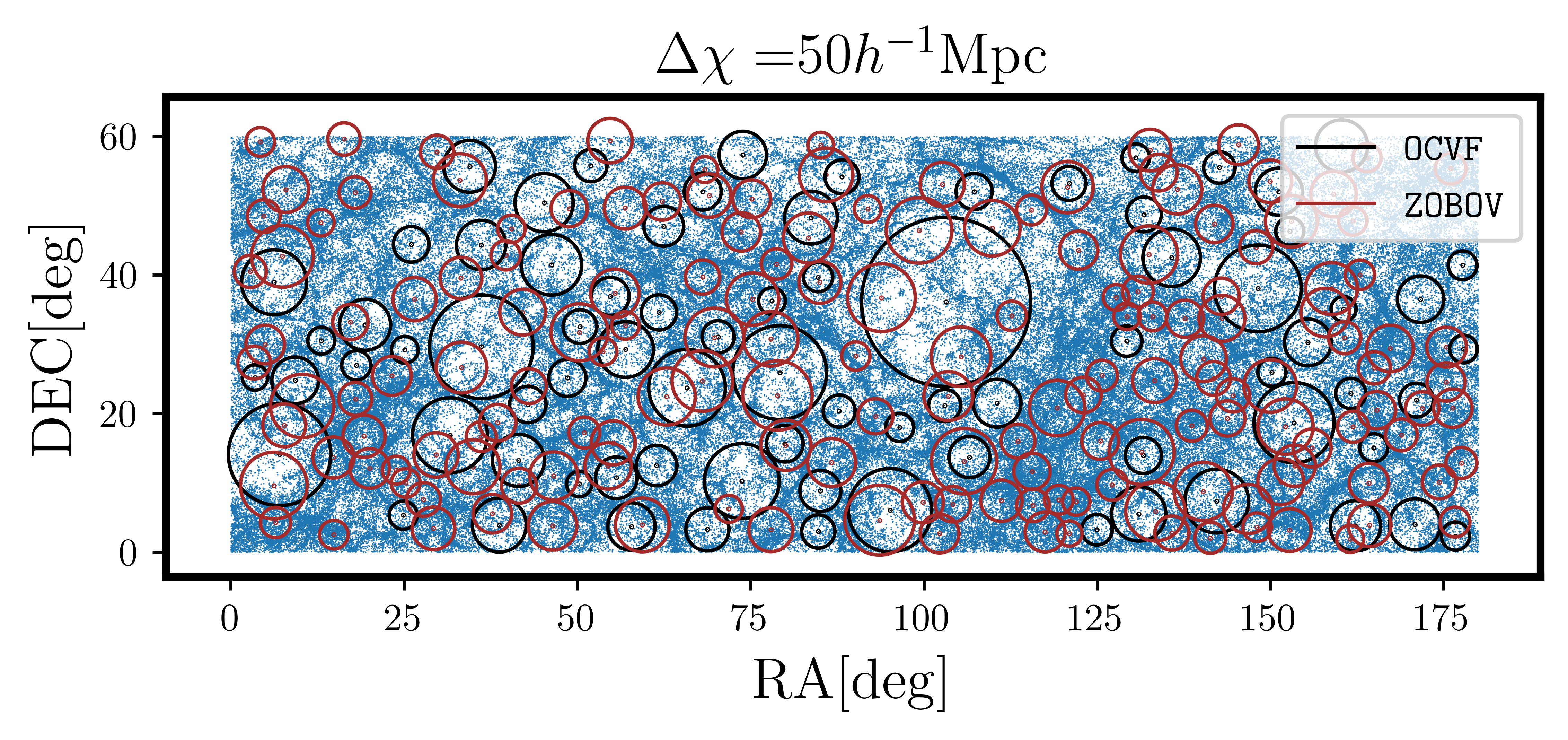}
\includegraphics[width=13.cm]{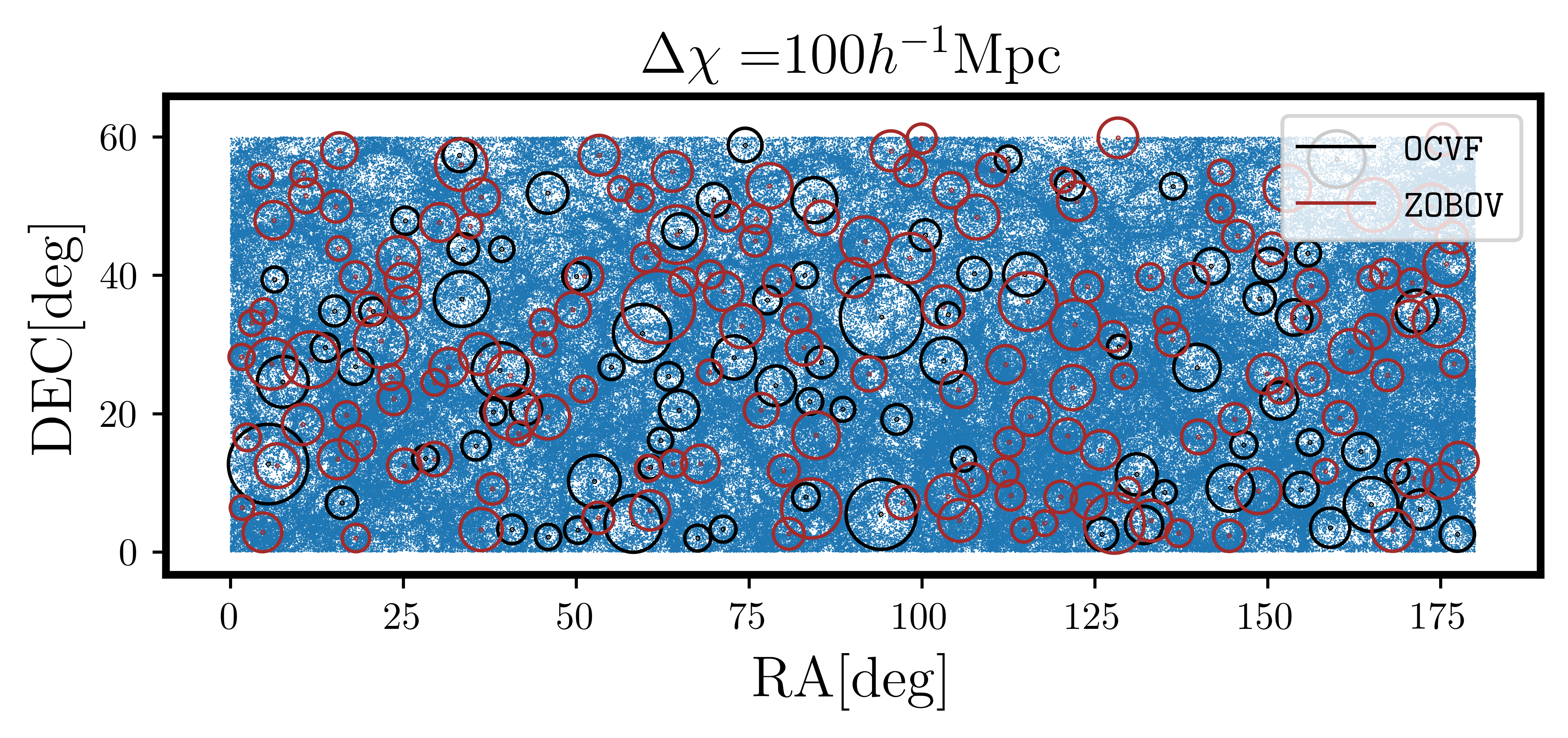}
\end{center}
\cprotect\caption{Visualisation of \verb|OCVF| and \verb|ZOBOV| voids plotted over BGS galaxies, in slices of widths $\Delta \chi = 20, $ $50$ and $100[h^{-1}\textrm{Mpc}]$. The \verb|ZOBOV| voids were chosen to have their central positions well within the slices.}
\label{voids_over_gals}
\end{figure}



We chose to show the comparison between the $2$D \verb|OCVF| because former works have indicated that voids found in projected fields are more efficient at measuring the weak-lensing signal by voids \cite{davies2021optimal} and more sensitive to modifications to gravity \cite{cautun2018santiago}. We confirmed that by using the same algorithm to find voids in the 2D and 3D fields, the former provides a signal with larger amplitude. The reason for this is that underdensities in the projected field correspond to anisotropic underdensities in the 3D field, with major-axis aligned with the line-of-sight, as shown in \ref{vl_dm}. As a consequence, the photon's path is guaranteed to be deviated by underdensities over a greater distance. In contrast, 3D voids are, in general, surrounded by overdensities, which partially or completely erase the signature of overdensities in the photon's path.

This comparison aims to highlight the qualitative differences between these two approaches, which evidences that unlike other statistics, the choice of void finder and how it is applied to data (in the 3D or 2D field, for instance) can drastically change the estimated observable ($\Delta \Sigma$ in this case). 

\subsection{Void catalogues}

 The \verb|OCVF| is applied to the light-cone by splitting it in slices of equal widths along the line of sight. We use the widths $\Delta \chi = 20, 50, 100h^{-1}\mathrm{Mpc}$. Therefore, voids are initially detected as underdensities in projected slices, which in fact correspond to voids in the 3D field as we will discuss in section \ref{vl_dm}. The \verb|ZOBOV| algorithm is applied to the $3$D galaxy field. 
 
\begin{figure}
\centering
\begin{minipage}[b]{0.45\textwidth}
\centering
\includegraphics[width=7.5cm]{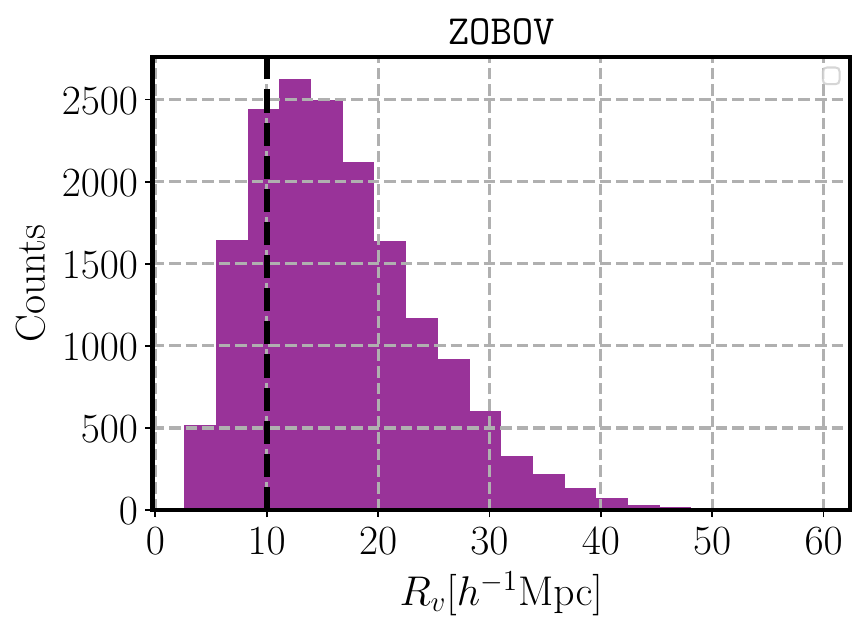} 
\end{minipage}\qquad
\begin{minipage}[b]{0.45\textwidth}
\centering
\includegraphics[width=7.5cm]{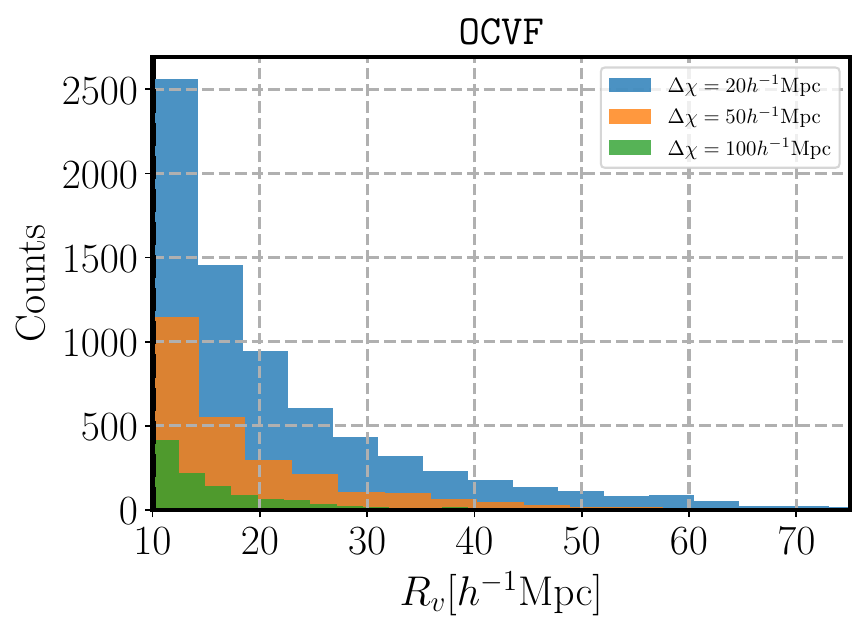} 
\end{minipage}
\cprotect\caption{Left: Radius distribution of \verb|ZOBOV| voids. Right: Radius distribution of \verb|OCVF| voids}
\label{radius_dist}
\end{figure}

 The void radius distributions in each version of \verb|OCVF| (slice width), as well as in \verb|ZOBOV| are shown in Figure \ref{radius_dist}. Table \ref{void_numbers} shows the total number of voids for each algorithm. We choose to apply a radius cut  $R_{v} > 10h^{-1}\mathrm{Mpc}$ in all void catalogues. This choice is related to the minimum resolution we have in the galaxy field, but is rather conservative, especially for the \verb|OCVF| sample, which has the vast majority of voids with radii 
 smaller than $10h^{-1}\mathrm{Mpc}$. However, we believe that these smaller voids are predominantly spurious, i.e. does not correspond to real underdensities in the DM field or are mostly voids-in-clouds. 
 
 The \verb|OCVF| voids found in larger slices are significantly less numerous than 
 those found in smaller slices, however, if we normalize the curves to be integrated to unity, all histograms have the same shape. 

Figure \ref{voids_over_gals} shows the visualisation of \verb|ZOBOV| and \verb|OCVF| voids plotted over the BGS galaxy distribution. The chosen \verb|ZOBOV| voids were those with central positions as close as possible from the bin center we used to find the \verb|OCVF| voids, in a way that these voids must correspond to the underdensities in this projected slice. By visual inspection it is possible to notice that the \verb|OCVF| voids are better fitted into the regions with less galaxies in the projected field, whereas \verb|ZOBOV| voids sometimes correspond to those regions but frequently are placed into overdense regions. This is expected, since \verb|OCVF| finds voids in the projected slices and \verb|ZOBOV| is applied to the $3$D field. It shows that three-dimensional voids are sometimes erased in the projected field and, as a consequence, won't be detectable through lensing, or will simply add noise to the estimated VL profile.

\subsection{The $\Delta \Sigma$ Estimator and the TL approximation}

The ESMD estimator is given by

\begin{equation}
\widehat{\Delta \Sigma} (r_{\perp}/\mathrm{R_{v}}) = \frac{\sum_{ij} w_{ij} \Sigma_{c, ij} \gamma_{t, ij}(r_{\perp}/R_{v}) }{\sum_{ij}w_{ij}},
\label{estimator}
\end{equation}
where $\gamma_{t,ij}$ is the tangential shear in the source galaxy $i$ due to the void $j$. For each pair, the weights that minimize the variance of the signal are $w_{ij} = \Sigma^{-2}_{c, ij}$ \cite{sheldon2004galaxy}, and the critical mass density is defined as 
\begin{equation}
\Sigma_{c, ij} = \frac{c^{2}}{4 \pi G} \frac{\chi (z_{i})}{\chi (z_{j})(\chi(z_{i}) - \chi(z_{j}))}.
\end{equation}

The tangential shear is calculated from the shear components $\gamma_{1,2}$ (defined in section \ref{sec:wl} and obtained in \cite{derose2019buzzard} through ray-tracing) and the angle $\phi$ between the void and source positions.

The estimated ESMD is expressed in terms of the reduced perpendicular distance to the line-of-sight $r_{\perp}/R_{v}$ because voids with different sizes present very similar profiles in reduced coordinates. This is what we mean by stacked void profile. 

The estimator \ref{estimator} is based under the assumption that the shear $\gamma_{t, ij}$ for \textit{all} voids can be obtained by assuming the thin lens approximation, in which the net effect on the photons path will depend only on the target structure which can be regarded as being contained in a single plane, i.e. ignoring the line-of-sight direction. In other words, the remaining structures when subtracting the void will have zero net effect on the photons path and the only effect can be computed by projecting the density field of the voids, as in eq. \ref{tl_def}. In this case the equality in \ref{estimator} is true. However, it is possible that individual voids extend to distances larger than the maximum distance which does not allow us to write that $\Delta \Sigma = \Sigma_c \gamma_{t}$. In this case, a systematic error will be introduced in the estimator \ref{estimator}.

\subsection{Void finder comparison}

\begin{figure}
\centering
\begin{minipage}[b]{0.45\textwidth}
\centering
\includegraphics[width=7.5cm]{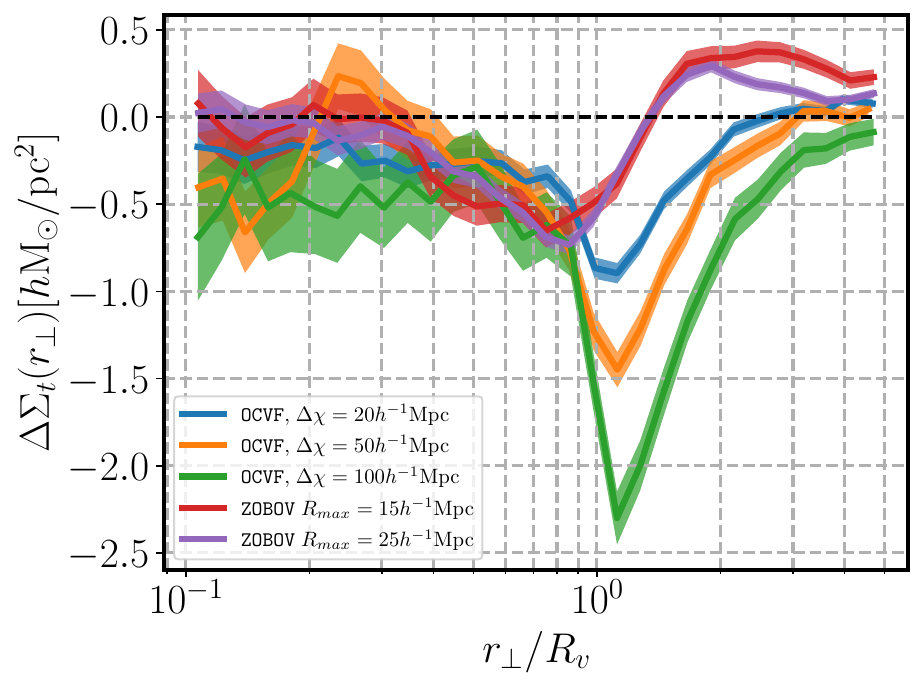} 
\end{minipage}\qquad
\begin{minipage}[b]{0.45\textwidth}
\centering
\includegraphics[width=7.5cm]{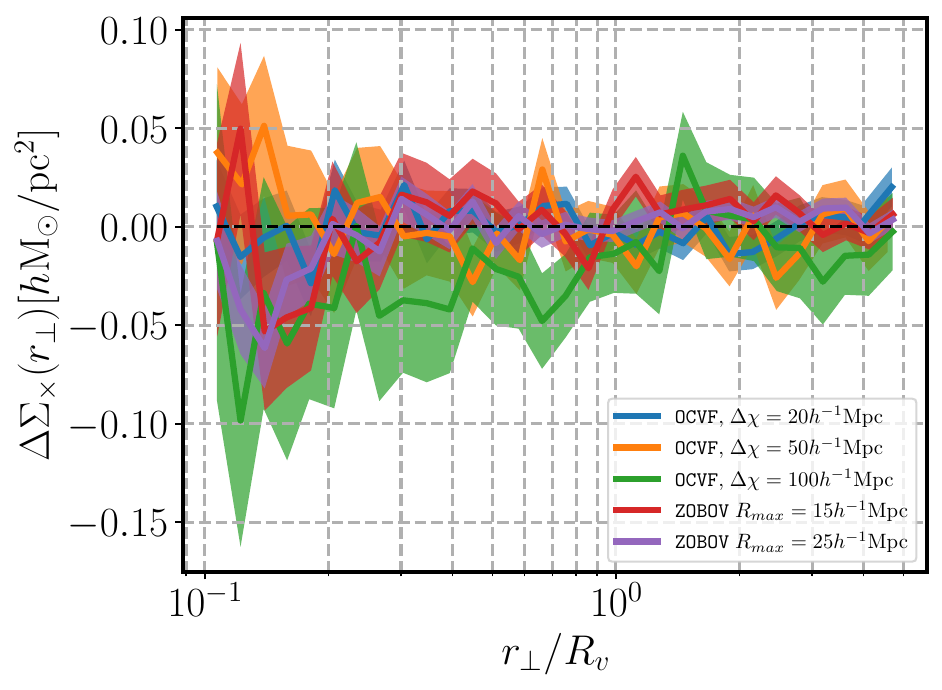} 
\end{minipage}
\cprotect\caption{Left: Comparison between the $\Delta \Sigma_{t}$ measurements performed using the \verb|OCVF| in slices of width $\Delta \chi = 20$, $50$, $100\,h^{-1}\,\mathrm{Mpc}$ (blue, orange, and green) and \verb|ZOBOV| in the bins of radius $[10, 15]h^{-1}\mathrm{Mpc}$ (red) and $[10, 25]h^{-1}\mathrm{Mpc}$ (purple). Right: The same comparison for the cross component $\Delta \Sigma_{\times}$.}
\label{comp_DS_t}
\end{figure}

In Figure \ref{comp_DS_t}, we show the measurement of the ESMD performed using voids found in the 3D-galaxy field with the \verb|ZOBOV| algorithm, as well as voids found in projected fields with the \verb|OCVF| algorithm. The \verb|ZOBOV| voids are split into two sub-samples, $R_{v} \in [10, 15] $ and $R_{v} \in [10, 25]$ ($h^{-1}\mathrm{Mpc}$). The reason why we use these two sub-samples and 25$h^{-1}\mathrm{Mpc}$ as maximum radius is justified in section \ref{sec:num_interp}. The \verb|OCVF| voids are presented in three samples, each of which corresponds to the algorithm applied to slices of widths $\Delta \chi = 20, 50, 100 h^{-1}\mathrm{Mpc}$. The right and left plots show, respectively, the tangential ($\Delta \Sigma_{t}$) and cross ($\Delta \Sigma_{\times}$) components.  

The tangential component presents several differences. Firstly, the \verb|ZOBOV| voids present a compensation wall, whereas the \verb|OCVF| does not. The compensation wall is a direct consequence of the same feature in the void's 3D density profile, which means that these voids are closer to overdensities, whereas \verb|OCVF| voids are far enough from overdensities to not  
 have correlation with them. Another notorious difference is that \verb|ZOBOV| voids are shallower. There are two reasons for this. The first is the existence of a compensation wall, which produces positive tangential shear, $\gamma_{t}$. The second reason can be intuitively seen in Figure \ref{voids_over_gals}: not all \verb|ZOBOV| voids correspond to underdensities in the projected field, whereas the voids found in projected fields are guaranteed 
to not 
 be correlated with overdensities along the line-of-sight. Moreover, larger projected slices tend to select underdensities aligned with the line-of-sight. The \verb|OCVF| voids are presented in three samples, each of which corresponds to the algorithm applied to slices of widths $\Delta \chi = 20, 50, 100 h^{-1}\mathrm{Mpc}$. These voids correspond to 3D underdense structures which have their major axis aligned with the line-of-sight (see Figure \ref{Profiles_2D}). In the Appendix \myrefB{appendix_b} we briefly discuss that not only the voids in the projected field present an anisotropic 3D density profile, but also the subsample voids found using the 3D version of the same algorithm (\verb|OCVF|), therefore following the theoretical abundance (Figure \ref{abund_sim}), also present anisotropic 3D density profiles. This subsample is chosen using the distances from the 3D voids to the projected voids which present correlation between them (Figure \ref{2d_3d_corr}). This result indicates that what we really find by applying the void finder in the projected field are combinations of 3D voids which present intrinsic alignment between them.

The cross-components must be consistent with zero. It is almost always satisfied, except in some cases for small $r_{\perp}/R_{v}$, where it is slightly bellow zero. The same feature also appears in the tangential component (it must go to zero as $r_{\perp}/R_{v} \rightarrow 0$). This indicates a systematic error for small $r_{\perp}/R_{v}$ that we do not comprehend.

\begin{figure}
\centering
\begin{minipage}[b]{0.45\textwidth}
\centering
\includegraphics[width=7.5cm]{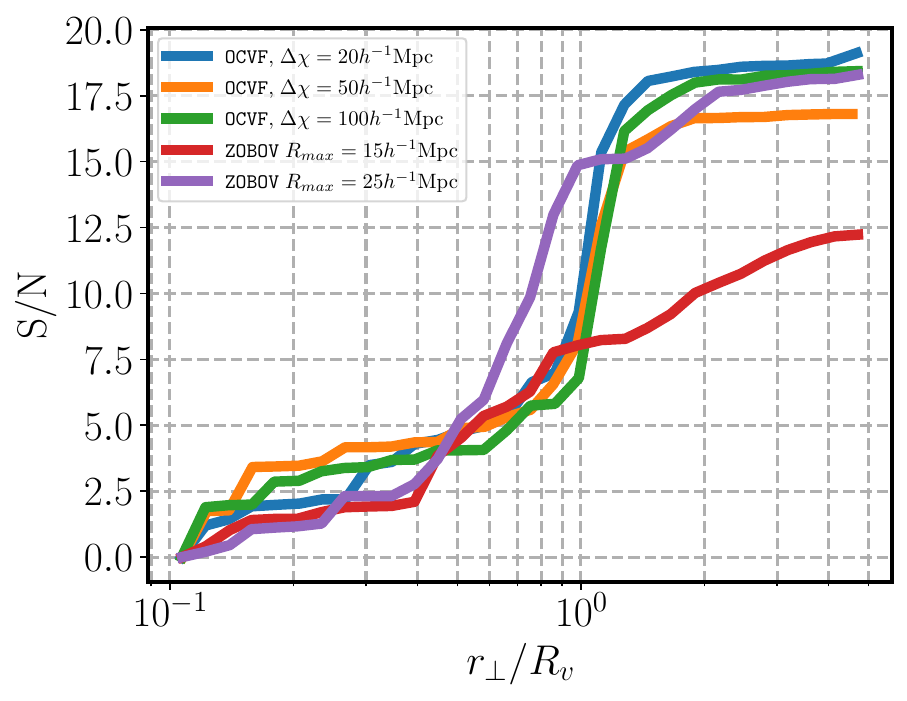} 
\end{minipage}\qquad
\begin{minipage}[b]{0.45\textwidth}
\centering
\includegraphics[width=7.5cm]{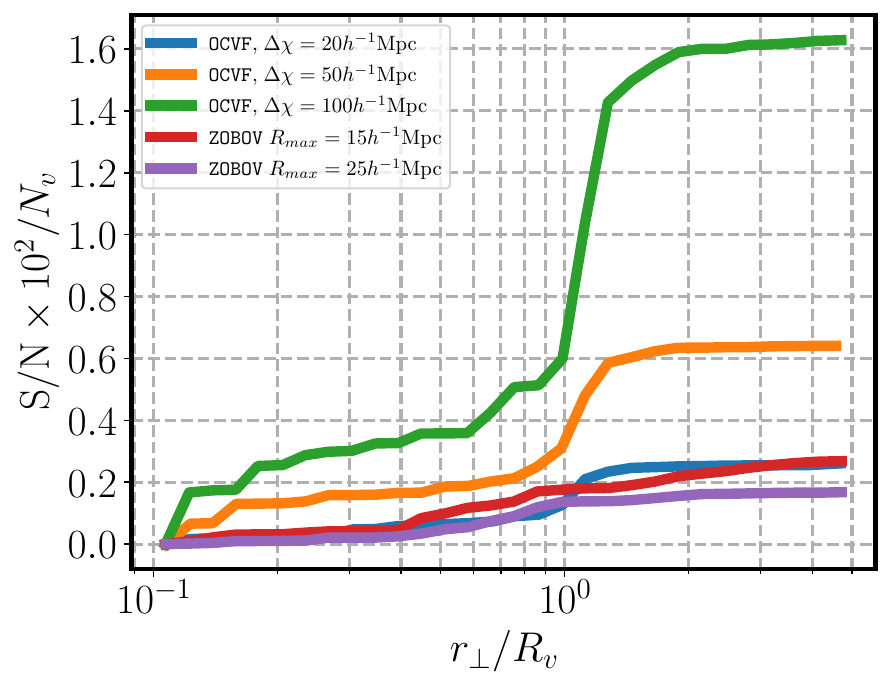} 
\end{minipage}
\cprotect\caption{Left: Comparison between the cumulative S/N on $\Delta \Sigma_{t}$ obtained using the \verb|OCVF| on slices of width $\Delta \chi = 20$, $50$, $100\,h^{-1}\,\mathrm{Mpc}$ (blue, orange, and green) and \verb|ZOBOV| (red and purple) in the redshift bin $0.1 < z < 0.3$. Right: S/N normalized by the number of voids}
\label{comp_SN}
\end{figure}

In Figure \ref{comp_SN} we compare the cumulative signal-to-noise (S/N) of the tangential component, defined as

\begin{equation}
(S/N)^{2}( < r_{\perp,k}) = \sum_{i \leq k, j \leq k} \Delta \Sigma_{t,i} C^{-1}_{ij} \Delta \Sigma_{t,j},
\end{equation}

\noindent where $C_{ij}$ is the covariance matrix estimated through a void-by-void jack-knife. 

The left plot shows the cumulative S/N for each sample, whereas the 
right plot shows the same quantity normalised by the number of voids in the corresponding sample (Table \ref{void_numbers}). The cumulative S/N is quite similar for all samples, except for the \verb|ZOBOV| sample with maximum radius of 15$h^{-1}\mathrm{Mpc}$. The normalised S/N is significantly higher for the \verb|OCVF| samples corresponding to the slices widths of $\Delta \chi = 50, 100h^{-1} \mathrm{Mpc}$. This result is expected: the larger is the slice width, the larger is the photon's path which corresponds to underdensities, i.e., less contamination from overdensities.

It should be stressed that this comparison does not assess which sample is more useful to constrain cosmology. Indeed, larger S/N does not necessarily mean more sensitive to cosmological parameters or modifications to gravity. However, we can say that measuring the lensing signal by voids using projected slices and the \verb|OCVF| will produce a signal which tells more about underdensities than the one obtained using \verb|ZOBOV| voids. By thinking of voids as being the most  underdense tracers of LSS and assuming that tracers with different biases carry complementary information (see e.g. \cite{mergulhao2022effective, abramo2013multitracer}), then the \verb|OCVF| profiles in Figure \ref{comp_DS_t} have biases more negative than the \verb|ZOBOV| profiles.  

\begin{figure}
\centering
\includegraphics[width=10cm]{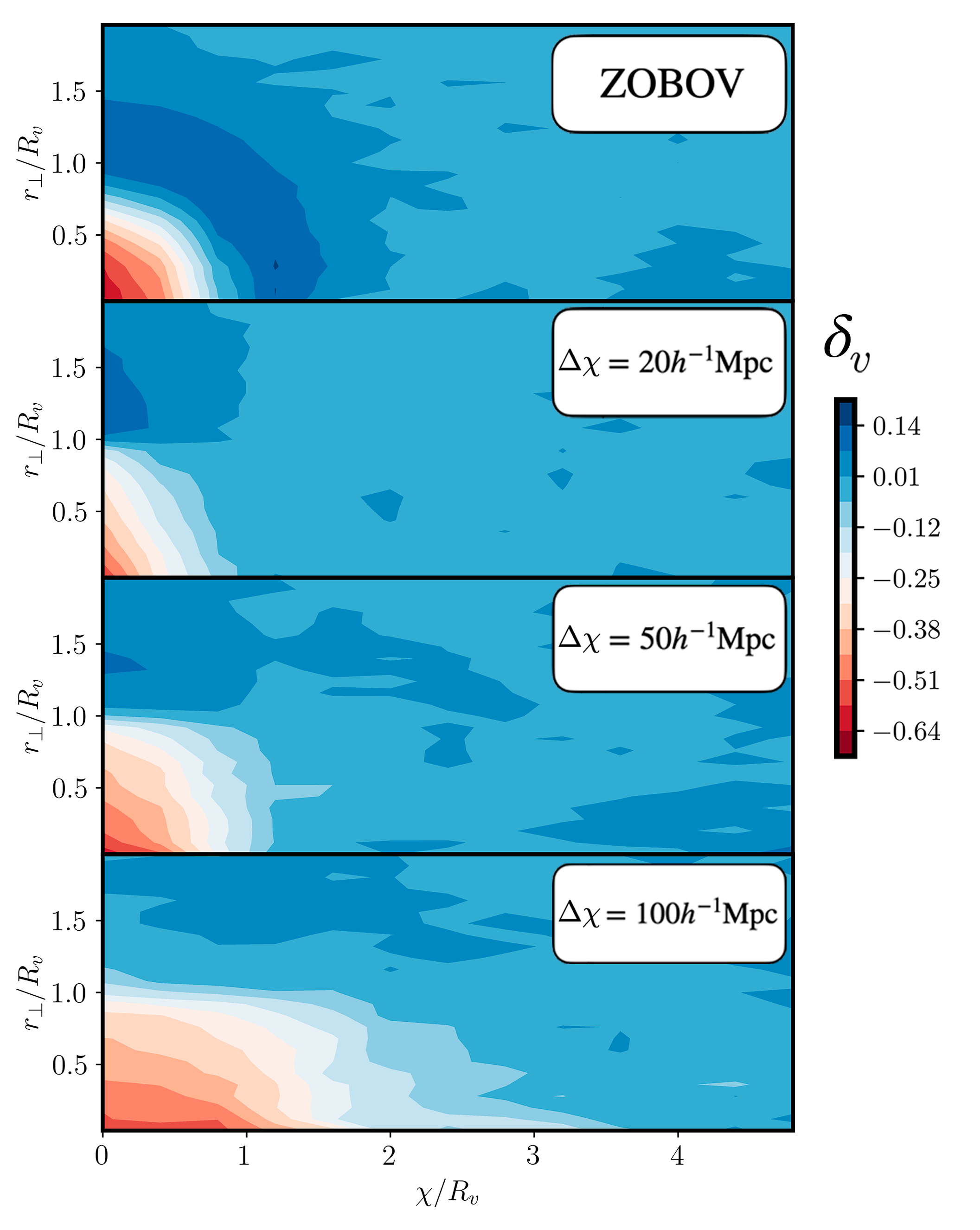} 
\cprotect\caption{Void DM profiles as a function of radial comoving distance, $\chi/R_{v}$ (centered at the void position) and the perpendicular distance to the line-of-sight, $r_{\perp}/R_{v}$. The top panel shows profiles estimated with \verb|ZOBOV| voids found in the 3D galaxy distribution, and the profiles labeled by $\Delta \chi = 20, 50, 100 h^{-1}\mathrm{Mpc}$ are \verb|OCVF| voids found in slices of labeled sizes.}
\label{Profiles_2D}
\end{figure}  
\begin{table}[]
 \centering
 \begin{tabular}{|c|c|c|}
 \hline
 Void Finder  & $ N_{v} (0.1 < z < 0.3) $  \\
 \hline 
 \verb|OCVF| ($\Delta \chi$ = 20 $h^{-1}\textrm{Mpc}$)   & 7321 \\
 \hline
 \verb|OCVF| ($\Delta \chi$ = 50 $h^{-1}\textrm{Mpc}$)   & 2623  \\
 \hline
 \verb|OCVF| ($\Delta \chi$ = 100 $h^{-1}\textrm{Mpc}$)  & 1131 \\
 \hline
 \verb|ZOBOV|   & 13438   \\
 \hline
 \end{tabular}
\cprotect\caption{The total number of voids for each void finder in the redshift bin $z$ $\in$ $[0.1, 0.3]$.}
 \label{void_numbers}
\end{table}



\section{Numerical interpretation}
\label{sec:num_interp}


Voids are peculiar tracers of LSS and it is not clear whether an analytic approach, based on perturbation theory, or effective field theory of large scale structure would be able to predict their density profiles, i.e., the cross-correlation void-tracer. The main reason for this impossibility is that voids are not uniquely defined, imposing a puzzle on how the condition related to the void center definition will be incorporated in the growth equation, in the case of linear perturbation theory. Therefore, a promising approach to extract this cosmological information is to take their profiles directly from N-body simulations and apply some emulation technique to predict their profiles for a set of cosmologies. In the context of VL, we have the advantage that we can directly use a DM only box to performe the emulation, since this observable is probing directly the total matter field. 

In this section we pave the way for this kind of approach. We use the DM particles of the same realisation we used in the previous section and estimate the DM profiles of the same voids we found in the galaxy field. Then we use these void DM profiles to check the consistency between the $\Delta \Sigma$ profiles we estimated in the previous section (through the shear of background galaxies), and the same quantity, but estimated directly from the DM field. This consistency test will show whether we are really having access to the DM profiles of these voids through weak-lensing. This consistency test is crucial if we aim to use the VL profile to do precision cosmology in the near future. As we will show, this consistency is not trivially given in the context of voids, as it is in the context of halos.

\subsection{VL and Void Dark Matter profiles}
\label{vl_dm}

Figure \ref{Profiles_2D} shows the stacked void profiles as a function of parallel and perpendicular distances to the line-of-sight, estimated using the DM particles of the Buzzard mock, i.e., the cross-correlation between void centers and DM particles for voids found in the BGS galaxy field:
\begin{equation}
\delta_{v}(r'_{\perp}, \chi') = \frac{1}{N_{v}}\sum^{N_{v}}_{i} \frac{n^{i}_{p}(\Delta r'_{\perp}, \Delta \chi')}{ \langle n_{p} \rangle  }  -1,
\label{prof_est}
\end{equation}
where $r'_{\perp} \equiv r_{\perp}/R_{v}$, $ \chi' \equiv \chi/R_{v}$ and $\Delta r'_{\perp}(\chi)$ denotes a bin in $(r'_{\perp}, \chi')$. 
Since we are estimating these profiles in configuration space, we expect that they will be isotropic. However, the profiles for increasingly larger slices become increasingly anisotropic. Since only voids aligned with the line-of-sight are detected as voids in the projected field, the resulting stacked profile will be anisotropic and might contain some information encoded in the void ellipticities of voids defined in the 3D distribution (see Appendix \ref{appendix_b}). This can be directly accessed with void lensing measurements. The relation between anisotropic and isotropic profiles is the topic of an ongoing work.

\begin{figure}
\centering
\includegraphics[width=0.7\textwidth]{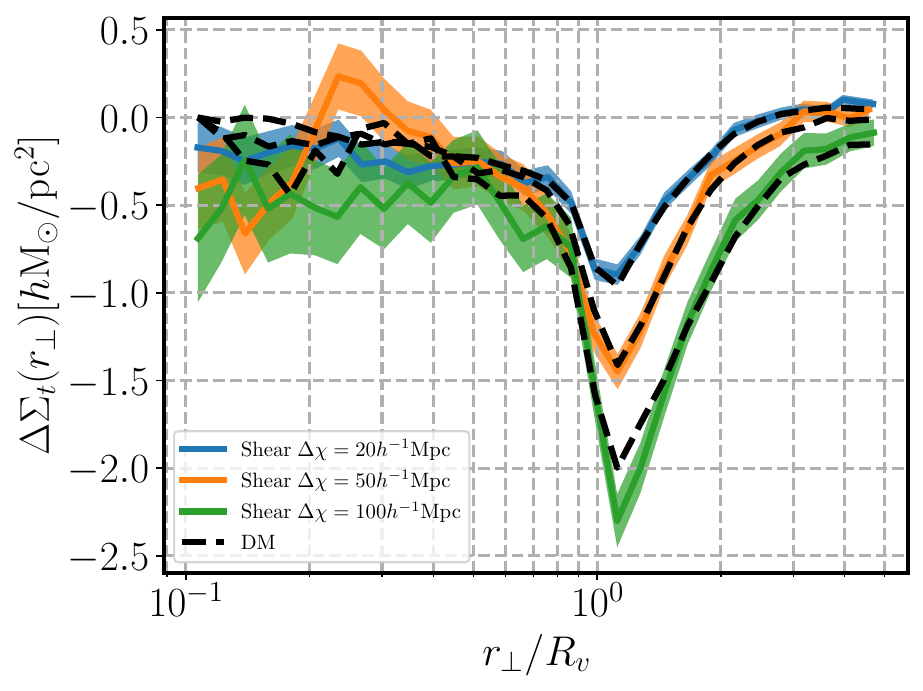} 
\caption{Comparison between $\Delta \Sigma_{t}$ as measured through the shear (blue, orange, and green) and directly using the DM particles (dashed black).}
\label{comp_shear_DM}
\end{figure}

\subsection{Consistency between shear and Dark matter density profile around voids}

In this section, we check the consistency between the $\Delta \Sigma_{t}$ profile of voids measured from the shear, i.e., $\Delta \Sigma_t= \Sigma_{c} \gamma_{t}$ and the one  directly calculated using the DM density profiles of voids presented in figure \ref{Profiles_2D}. We take the same realisation of the density field, as traced by BGS galaxies. Therefore, we compare the observable we have access through observations against the same observable but computed using the DM field, which we do not have access to in real observations. In the case where this consistency test is successful, we know exactly what we would observe in real data, given that the assumption we are using in simulations are well calibrated w.r.t real data, namely, the galaxy bias and possible systematic effects such as intrinsic alignment and survey mask.
Before proceeding, it is important to mention the reasons why the two quantities might differ:

\begin{itemize}
    \item \underline{The impact of the source distribution:} In the estimator \ref{estimator}, the weights and the $\gamma_t$ depend on $\Sigma_{c}$ which contains information about the source distribution. By predicting $\Delta \Sigma_t$ only in terms of the void DM profile, without saying anything about the source distribution, it is not clear whether we can recover the same quantity than the one estimated.
    \item \underline{The impact of the thin lens approximation:} To quote the results in terms of $\Delta \Sigma_t$ we are assuming that the void profile, or everything that will have an impact on the average shape of background galaxies is contained in a thin lens (see eq. \ref{tl_def}). Since voids can have radius as large as $\simeq 100h^{-1}$Mpc and, furthermore, might have correlations with overdensities beyond the void radius, it is not clear whether the thin lens assumption still holds.
\end{itemize} 

In order to calculate the $\Delta \Sigma_t$ profiles from the 3D void profiles in figure \ref{Profiles_2D}, we will exclude small scales to avoid resolution effects by using the annular differential surface density (ADSD):

\begin{equation}
\begin{aligned}
\Upsilon\left(R | R_0\right) \equiv & \Delta \Sigma_t(R)-\frac{R_0^2}{R^2} \Delta \Sigma_t\left(R_0\right) \\
= & \frac{2}{R^2} \int_{R_0}^R \mathrm{~d} R^{\prime} R^{\prime} \Sigma\left(R^{\prime}\right)-\frac{1}{R^2}\left[R^2 \Sigma(R)-R_0^2 \Sigma\left(R_0\right)\right],
\end{aligned}
\label{ADSD}
\end{equation}

\noindent where $R_{0}$ is the cutting scale. For sufficiently small $R_{0}$, $\Upsilon$ reduces to $\Delta \Sigma_t$. We checked that the estimated profile does not depend on the particular choice of $R_{0}$, for $R_{0} \leq 0.5 $ in reduced coordinate ($r_{\perp}/R_{v}$). 

Since we are working with stacked void profiles, $\Delta \Sigma$ is proportional to the void radius, which comes from the Jacobian when tranforming the integral from $r'_{\perp}$ to $r_{\perp}/R_{v}$ coordinate. Thus,

\begin{equation}
\Sigma (r'_{\perp}) = R_{v} \Sigma(r_{\perp}),
\end{equation}
where $\Sigma (r_{\perp})$ is given by eq. \ref{Sigma_def} and, consequently, 

\begin{equation}
\Delta \Sigma_t(r'_{\perp}) = R_{v} \Delta \Sigma_t (r_{\perp}).
\end{equation}

\begin{figure}
\centering
\begin{minipage}[b]{0.45\textwidth}
\centering
\includegraphics[width=7.5cm]{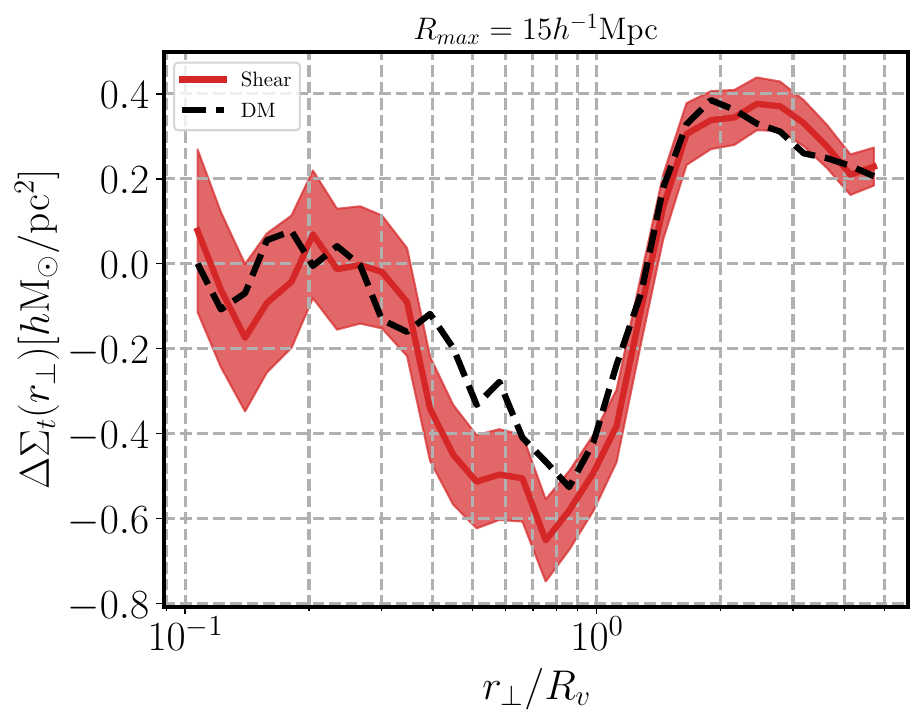} 
\end{minipage}\qquad
\begin{minipage}[b]{0.45\textwidth}
\centering
\includegraphics[width=7.5cm]{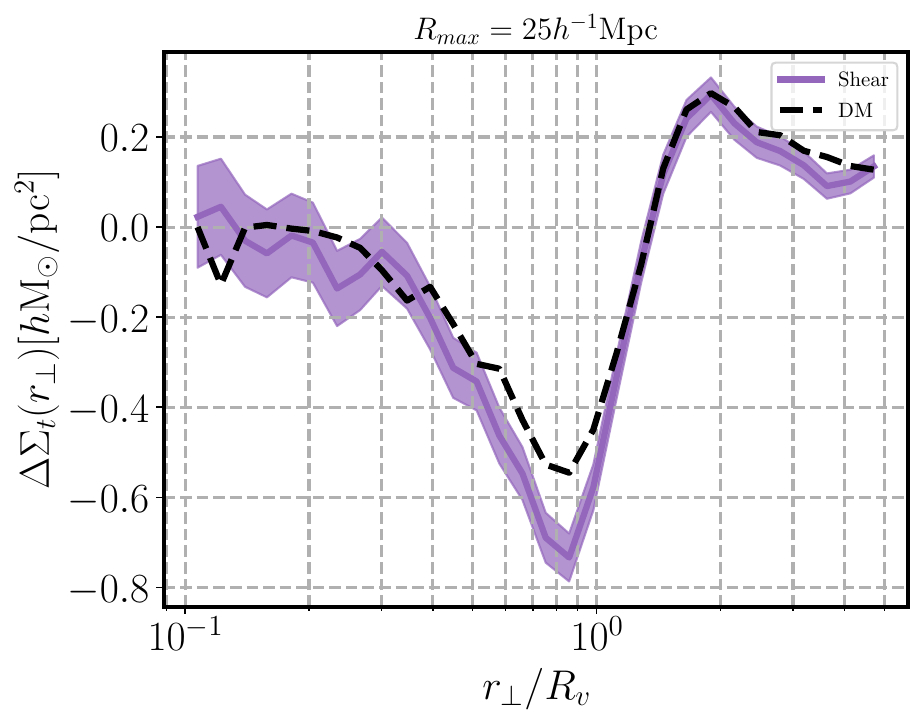} 
\end{minipage}
\begin{minipage}[b]{0.45\textwidth}
\centering
\includegraphics[width=7.5cm]{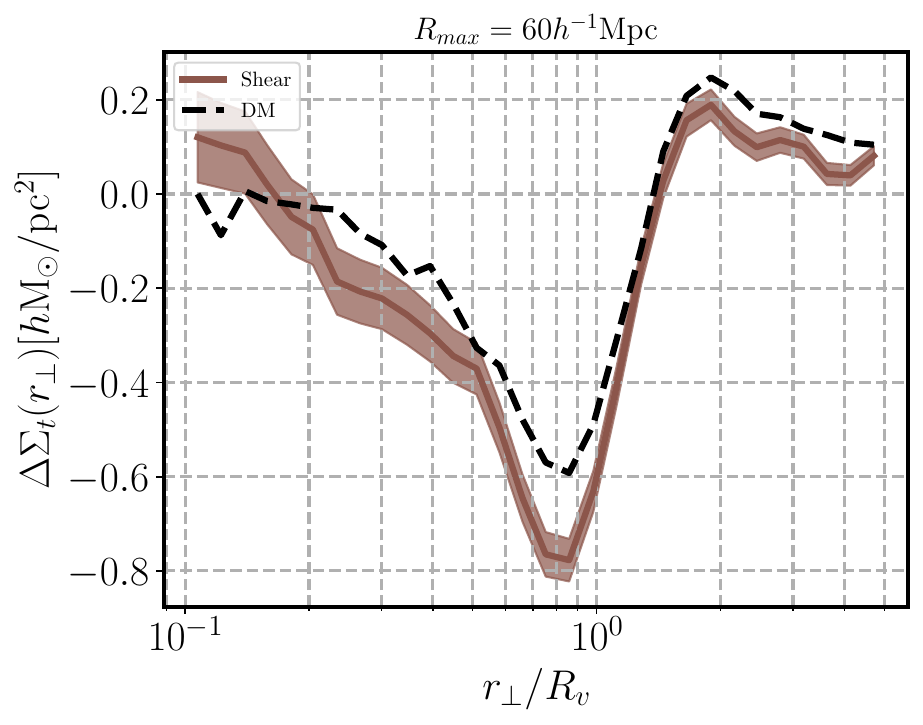} 
\end{minipage}

\cprotect\caption{The consistency check between the $\Delta \Sigma_{t}$ estimated through the shear of background galaxies and the same quantity estimated directly from the DM profiles of the same voids. Respectively, the top-left, top-right and bottom plots show the consistency test for \verb|ZOBOV| voids with radii in the ranges $[10, 15]$, $[10, 25]$ and $[10, 60]$ in units of $h^{-1}\mathrm{Mpc}$.   }
\label{comp_shear_DM_rev}
\end{figure}

Figure \ref{comp_shear_DM} shows the results of the same $\Delta \Sigma_{t}$ profiles as shown in figure \ref{comp_DS_t}, through the estimator $\ref{estimator}$, against the $\Delta \Sigma_{t}$ estimated using the DM particles around the voids. In this estimation we project the void profiles using a bin width $\Delta \chi'$:

\begin{equation}
\Sigma (r_{\perp}) = R_{v}\int_{\chi'_{l} - \Delta \chi'/2.}^{\chi'_{l} + \Delta \chi'/2.} \frac{d\chi'}{a(\chi')}\delta_{v}(r'_{\perp}, \chi').
\end{equation}

\noindent and then compute the ADSD signal using eq. \ref{ADSD}.

We obtain consistency between shear and DM profiles for the \verb|OCVF| samples computed in slices of widths $\Delta \chi = 20, 50h^{-1}\mathrm{Mpc}$. However for the sample with $L = 100h^{-1}\mathrm{Mpc}$, the two profiles become inconsistent at reduced radius $r'_{\perp} \sim 1$. We obtain the same trend for the \verb|ZOBOV| voids (Figure \ref{comp_shear_DM_rev}). The subsample in the radius bin [10, 15] ($h^{-1}\mathrm{Mpc}$) is quite consistent, whereas the subsamples including larger voids, namely, [10, 25]($h^{-1}\mathrm{Mpc}$) and [10, 60]($h^{-1}\mathrm{Mpc}$), present increasing inconsistencies.

The trend with both void finders is, the larger is the stacked void profile, the larger is the inconsistency between shear and DM profiles. Furthermore, the void size in the direction perpendicular to the line-of-sight seems to be unimportant, since the normalised radius distribution of \verb|OCVF| voids is the same amongst the different bins widths. We conclude that what really matters for this consistency is the void size along the line-of-sight, which indicates that the thin lens approximation might be broken in the cases in which the correlations between the void and its surroundings extend beyond a certain limit. 

In the Appendix \ref{appendix_a} we show an example of how a break of the thin lens approximation affects the VL profiles. We use generic analytical void profiles and calculate $\Delta \Sigma$ with and without the thin lens approximation. The wrong assumption of the thin lens approximation tends to underestimate the VL profile, specially around $r_{\perp}/R_{v} \simeq 1$, which is exatly what we see in the DM profiles of Figures \ref{comp_shear_DM} and \ref{comp_shear_DM_rev}. The source of the inconsistency might be, then, the wrong assumption of the thin lens approximation in the estimator \ref{estimator}.

\section{Conclusions}

In this work we have studied the VL profile in the context of galaxy mock. First, we proposed a new void finder algorithm which is particularly designed to deliver voids with deep VL profiles. Then, we apply this algorithm to the galaxy mock and contrast it with the widely used algorithm in literature, \verb|ZOBOV|. The latter has been used in previous measurements of the VL profile in real data and provided a higher S/N compared to an algorithm similar to ours \cite{fang2019dark}. 

We show that, compared to \verb|ZOBOV|, voids found in projected slices using our algorithm can provide a higher S/N per void, with a deeper VL profile. These voids correspond to combinations of voids in the 3D DM field which present intrinsic alignment between them, suggesting that the VL profile might be sensitive to the void intrinsic alignment. This opens some questions, namely: (i) is the void intrinsic alignment connected to tidal fields in LSS and, therefore, to cosmology? (ii) How sensitive is the VL profile to the intrinsic alignment? 

Then we checked the consistency between the VL profile as estimated through the shear of background galaxies and the same quantity as estimated directly from the DM density profiles of voids. This consistency test has never been made (up to the knowledge of the present authors) and it is crucial for future cosmological analysis using VL profile as an observable. Unlike halos, voids have density profiles that extend 
over hundreds of $h^{-1}\mathrm{Mpc}$, which might break the assumption of voids being contained in a thin slice, which is a basic assumption when estimating the VL profile from the shear of background galaxies. Furthermore, voids are much less numerous than halos and therefore residual contributions from structures along the line-of-sight might not be averaged out, as it easily is in the context of halos. 

This consistency test shows that voids with larger sizes along the line-of-sight present inconsistencies between the shear and DM VL lensing profiles, suggesting that the thin lens approximation assumed in the estimator is not appropriate in the case of voids in general. For \verb|ZOBOV| voids smaller than $15h^{-1}\mathrm{Mpc}$ and \verb|OCVF| found in projected slices smaller than $50h^{-1}\mathrm{Mpc}$ the shear and DM VL profiles are consistent. 

In future works, we plan to further understand the relation between the VL profile around voids in the projected field and the intrinsic alignments of voids in the 3D field. Also, we plan to model the VL profile with dependency on the cosmological parameters. This work paves the way for a trivial numerical prediction: since we have shown that for a suitable choice of void definition, the shear and DM VL profile are consistent, emulation techniques which are becoming widely used in cosmology can be used to interpret real data observations. Is it also important to have an analytical prediction of the VL observable, which is the subject of an ongoing work by the present author.

\section*{Acknowledgements}

RB would like to thank Chris Blake and Joe DeRose for kindly making the Buzzard galaxy mock available, and Rodrigo Voivodic, Beatriz Tucci and Francisco Maion for helpful discussions in the preparation of this work. 
We acknowledge the financial support by the Excellence Initiative of Aix-Marseille University -A*MIDEX (AMX-19-IET-008 -IPhU), part of the France 2030 investment plan.

\appendix

\section{How the thin lens approximation affects the VL profiles}
\label{appendix_a}

In this appendix we explore the limits of the thin lens approximation, which is relevant in the case of voids, given their large size. 

As pointed out in section \ref{sec:num_interp}, the matching between measured and predicted ESMD depends on whether we can consider voids as thin lenses in between the source and the observer or not, i.e., whether we can write the convergence as eq.~\ref{tl_def}. To test this assumption we use a void profile with similar shape to the stacked profile produced by \verb|OCVF| to compute $\Delta \Sigma$ with and without the thin lens approximation. We use a ``worst case scenario'' in which the void position is too close to the source with redshifts $z_{l} = 0.48$ and $z_{s} = 0.5$, respectively for the void and the source, and vary the void radius $R_{v} = (50, 100, 150)h^{-1}\mathrm{Mpc}$. The combination of void radius and distance between the source and the void are the variables that control whether the thin lens is a good approximation or not.

\begin{figure}
\centering
\begin{minipage}[b]{.4\textwidth}
\includegraphics[width=7cm]{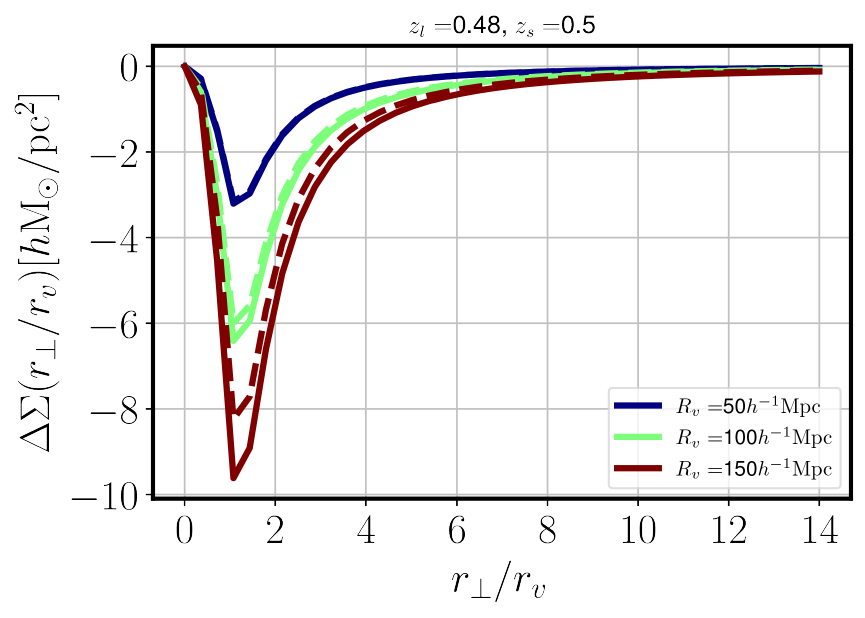} 
\end{minipage}\qquad
\begin{minipage}[b]{.4\textwidth}
\includegraphics[width=7cm]{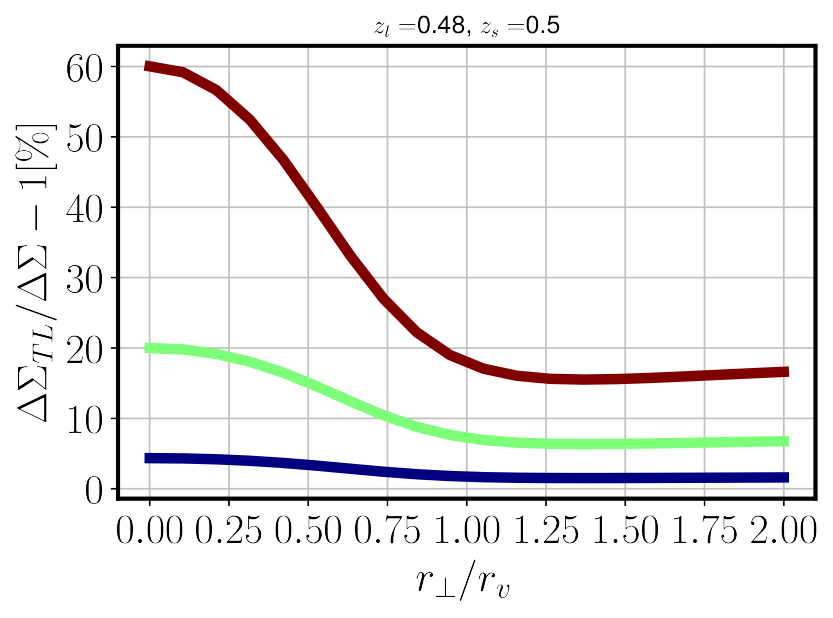} 
\end{minipage}
\caption{ Left: The ESMD profiles for different void radius with (solid) and without (dashed) the thin lens approximation. Right: The relative difference between the ESMD with and without thin lens approximation. }
\label{tl_app}
\end{figure}

Figure \ref{tl_app} shows the predictions for the ESMD for different void radius (left) and the relative difference between the same quantity calculated with and without the thin lens approximation (right). This exercise shows that for voids with radius of $50h^{-1}\mathrm{Mpc}$ the difference is below $\simeq 5 \%$.

\section{ The void intrinsic alignment around projected voids}
\label{appendix_b}

In this appendix we show some preliminary results about the shapes of 3D voids around the voids found in the projected slices that we use in this work. Both void samples were created using the \verb|OCVF| algorithm. 

\begin{figure}
\centering
\begin{minipage}[b]{.4\textwidth}
\includegraphics[width=7cm]{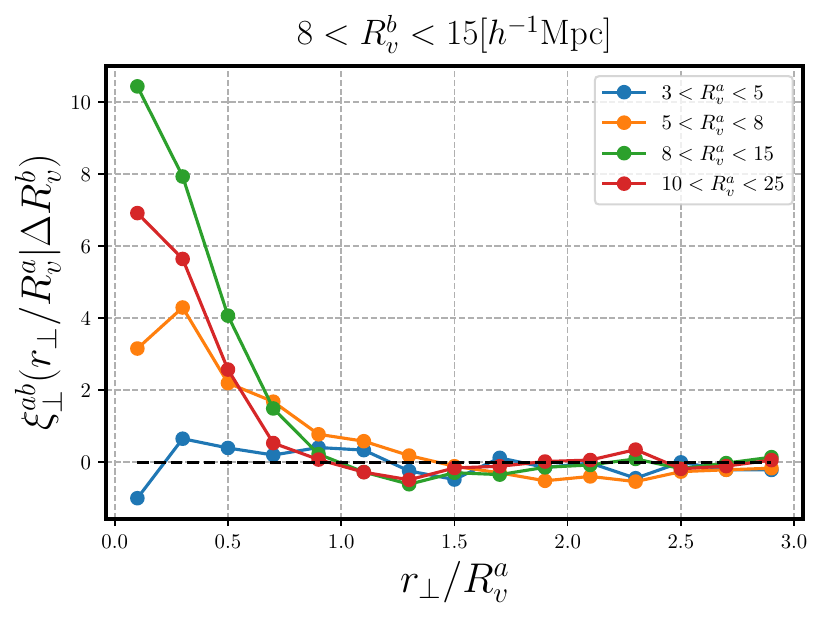} 
\end{minipage}\qquad
\begin{minipage}[b]{.4\textwidth}
\includegraphics[width=7cm]{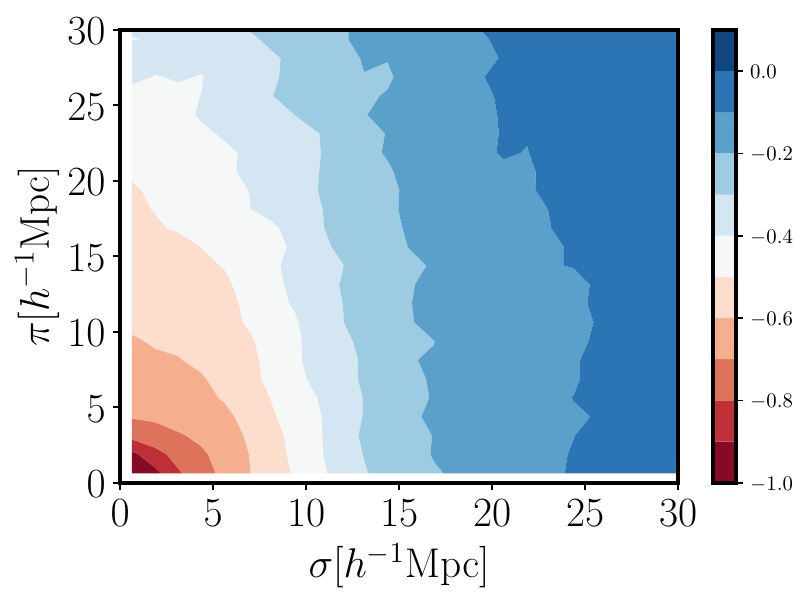} 
\end{minipage}
\caption{Left: Correlation between voids found in the projected field, labelled by ``b'' and voids found in the 3D field, labelled by ``a''. Right: the stacked profile of 3D voids which are at a distance from the projected voids which presents non-zero correlation, as a function of the parallel, $\pi$, and perpendicular, $\sigma$, distance to the line-of-sight.}
\label{2d_3d_corr}
\end{figure}

We use the same simulation we used in section \ref{sec:pheno} to find both void samples. The left plot in Figure \ref{2d_3d_corr} shows the correlation between a sample of voids found in projected fields of width $\Delta \pi = 50h^{-1}\mathrm{Mpc}$ with radii in the range $8 < R^{b}_{v} < 15h^{-1}\mathrm{Mpc}$ and the 3D voids in different radius bins. The right plot shows the stacked profile of the 3D voids as a function of the perpendicular, $\sigma$, and parallel, $\pi$, distances to the line of sight. 

The correlations shows that voids in $3$D and $2$D with similar sizes present stronger correlations. The stacked profiles of the $3$D voids are clearly anisotropic. This shows that $3$D voids around voids found in the projected slices present intrinsic alignment between them, or, equivalently, that the voids found in projected slices are actually the combination of $3$D voids, which are aligned between them.

A more detailed study of the correlations and the statistical relation between voids found in projected slices and $3$D voids is the subject of ongoing work.

\bibliographystyle{JHEP}
\bibliography{main}

\end{document}